\documentclass[twocolumn,twocolappendix,tighten]{aastex631}

\usepackage[flushleft]{threeparttable}
\usepackage{multirow}
\usepackage{enumitem}
\usepackage{newtxtext,newtxmath,ulem}
\usepackage[T1]{fontenc}
\usepackage{amsmath}
\usepackage{xcolor}

\def\be{\begin{equation}}
\def\ee{\end{equation}}
\def\ba{\begin{eqnarray}}
\def\ea{\end{eqnarray}}

\def\ltsima{$\; \buildrel < \over \sim \;$}
\def\simlt{\lower.5ex\hbox{\ltsima}}
\def\gtsima{$\; \buildrel > \over \sim \;$}
\def\simgt{\lower.5ex\hbox{\gtsima}}

\definecolor{falured}{rgb}{0.5, 0.09, 0.09}

\def\cv{{{\rm C}\,{\sc v}~}}
\def\cvi{{{\rm C}\,{\sc vi}~}}
\def\nvi{{{\rm N}\,{\sc vi}~}}
\def\nvii{{{\rm N}\,{\sc vii}~}}
\def\oiv{{{\rm O}\,{\sc iv}~}}
\def\ov{{{\rm O}\,{\sc v}~}}
\def\ovii{{{\rm O}\,{\sc vii}~}}
\def\oviii{{{\rm O}\,{\sc viii}~}}
\def\neix{{{\rm Ne}\,{\sc ix}~}}
\def\nex{{{\rm Ne}\,{\sc x}~}}
\def\mgxi{{{\rm Mg}\,{\sc xi}~}}
\def\mgxii{{{\rm Mg}\,{\sc xii}~}}
\def\sixiii{{{\rm Si}\,{\sc xiii}~}}
\def\sixiv{{{\rm Si}\,{\sc xiv}~}}
\def\sxvi{{{\rm S}\,{\sc xvi}~}}
\def\mgii{{{\rm Mg}\,{\sc ii}~}}
\def\cii{{{\rm C}\,{\sc ii}~}}
\def\caii{{{\rm Ca}\,{\sc ii}~}}
\def\civ{{{\rm C}\,{\sc iv}~}}
\def\siii{{{\rm Si}\,{\sc ii}~}}
\def\siiv{{{\rm Si}\,{\sc iv}~}}
\def\ovi{{{\rm O}\,{\sc vi}~}}
\def\oii{{{\rm O}\,{\sc ii}~}}

\shorttitle{Super-virial gas towards PKS\,2155-304}
\shortauthors{Bisht et al.}

\begin{document}

\title{Detection of `super-virial' gas in the Circumgalactic medium of the Milky Way towards PKS\,2155-304}

\correspondingauthor{Mukesh Singh Bisht}
\email{msbisht@rrimail.rri.res.in}

\author[0000-0002-1497-4645]{Mukesh Singh Bisht}
\affiliation{Raman Research Institute, Bangalore 560080, India} 

\author[0000-0002-9069-7061]{Sanskriti Das}
\altaffiliation{Hubble Fellow}
\affil{Kavli Institute for Particle Astrophysics and Cosmology, Stanford University, 452 Lomita Mall, Stanford, CA\,94305, USA}

\author[0000-0002-4822-3559]{Smita Mathur}
\affiliation{Astronomy Department, Ohio State University, 140 West 18th Avenue, Columbus, OH 43210, USA}
\affiliation{Center for Cosmology and Astro-particle Physics, Ohio State University, 191 West Woodruff Avenue, Columbus, OH 43210, USA}

\author[0000-0001-9567-8807]{Manami Roy}
\affiliation{Center for Cosmology and Astro-particle Physics, Ohio State University, 191 West Woodruff Avenue, Columbus, OH 43210, USA}
\affiliation{Astronomy Department, Ohio State University, 140 West 18th Avenue, Columbus, OH 43210, USA}

\author[0000-0001-6291-5239]{Yair Krongold}
\affiliation{Instituto de Astronomia, Universidad Nacional Autonoma de Mexico, 04510 Mexico City, Mexico}

\author[0000-0003-1880-1474]{Anjali Gupta}
\affil{Columbus State Community College, 550 E Spring St., Columbus, OH 43210, USA}

\begin{abstract}

We present the first simultaneous detection of four distinct highly ionized $z=0$ absorbing phases using Chandra and XMM-Newton grating spectra toward the blazar PKS\,2155–304. We detect the \mgxii K$\alpha$ absorption line for the first time in the circumgalactic medium (CGM) of the Milky Way. Along with \mgxii K$\alpha$, we detect \sixiv K$\alpha$ absorption, which are the tell-tale signatures of the hot `super-virial' gas in the CGM. Both from the model-independent calculations and hybrid-ionization modeling, we infer four phases at distinct temperatures, hot `super-virial' ($5.4^{+1.9}_{-0.8} \times 10^7$ K), warm-hot `virial' ($1.8^{+0.3}_{-0.2} \times 10^6$ K), warm `sub-virial' ($2.2\pm 0.5 \times 10^5$ K), and cool phase ($<1.7 \times 10^5$ K). The warm-hot and hot phases are $\alpha$-enhanced, and [C/O] and [Ne/O] are super-solar in the warm-hot phase, while [Mg/O] and [Si/O] are super-solar in the hot phase. The low-ionization lines are blue-shifted (v$_{\rm los} \approx -100$ km s$^{-1}$), whereas the high-ionization lines are red-shifted. It suggests a scenario of infalling sub-virial, quasi-static virial, and outflowing super-virial phases along this sightline. Earlier studies on individual sightlines were confined to the Northern Hemisphere. Our sightline is located in the Southern hemisphere, demonstrating that hot super-virial gas is also present at Southern Galactic latitudes as well. This confirms a more widespread distribution of the super-virial gas across both hemispheres.

\end{abstract}



\section{Introduction} 
\label{sec:introduction}

Galaxy disks are surrounded by the vast reservoir of gas known as the Circumgalactic medium (CGM; \citealt{Tumlinson2017}). The circumgalactic gas plays a vital role in galactic evolution, as it includes both inflowing material from the intergalactic medium (IGM), which serves as fuel for star formation, and outflowing gas expelled by stellar feedback and active galactic nucleus (AGN) activity. Depending on the physical processes at play in the CGM (\citealt{Donahue2022, Faucher2023}), the gas may either remain in the halo for extended periods due to long cooling times or cool and condense, eventually falling back onto the galactic disk to fuel future star formation.

The CGM of the Milky Way (MW) is diffuse and multiphase with temperatures ranging from the cool ($\sim 10^4$ K), warm ($\sim 10^5$ K), to the warm-hot phase at the virial temperature ($\sim 10^6$ K) and number density varying by $2-3$ orders of magnitude $\sim 10^{-4}-10^{-1} \, \rm cm^{-3}$. The most massive and the most volume-filling phase is the warm-hot phase (\citealt{Mathur2022}). The cool and warm gas in the CGM is typically detected as UV/optical absorption in the spectra of the bright background sources like quasars using tracers like \mgii, \cii, \caii, \siii, \siiv, \civ, \ovi, etc. The warm-hot virial phase is detected in both X-ray absorption (\citealt{Nicastro2002,Williams2005,Gupta2012,Gupta2014,Nicastro2016,Gupta2017,Gatuzz2018}) and X-ray emission (\citealt{Kuntz2000, Yoshino2009,Henley2010,Henley2013,Henley2015,Nakashima2018,Kaaret2020}). 

Recently, even hotter ($\sim 10^7$ K) `super-virial' gas has been detected in the CGM of the MW coexisting with the virial phase. 
This hot gas has been found in X-ray absorption (\citealt{Das2019a,Das2021,Lara-DI2023,McClain2023,Lara-DI2024a}) and X-ray emission (\citealt{Das2019b,Gupta2021,Bluem2022,Bhattacharyya2023,Ponti2023,Sugiyama2023,Gupta2023}). 
By studying and analyzing the X-ray data toward the Galactic X-ray Binaries (XRBs), \citealt{Lara-DI2024b} and \citealt{Roy2025} concluded that indeed the hot gas or the `super-virial' gas detected in absorption has an origin beyond the Galactic disk, from the extra-planar region or the extended medium. Using `shadow observations' with \textit{Suzaku}, \citealt{Gupta2025} conclusively showed that the hot gas in emission is also from regions beyond the Galactic disk. However, the temperature and chemical composition in emission differ from those in absorption; thus, the super-virial temperature gas detected in absorption and emission is not the same phase, even though both the phases may exist in the extra-planar region of the Galaxy. 

In X-ray absorption, the super-virial gas is detected along three individual sightlines with high quality grating spectra, 1ES\,1553+113 (\citealt{Das2019a}; $l=21.21^{\circ},b=43.96^{\circ}$), Mrk\,421 (\citealt{Das2021}; $l=179.83^{\circ},b=65.03^{\circ}$), and NGC\,3783 (\citealt{McClain2023}; $l=287.46^{\circ},b=22.95^{\circ}$) using the absorption lines of Hydrogen-like and Helium-like ions of Carbon, Nitrogen, Oxygen, Neon, Magnesium, and Silicon in the continuum of the background quasar. These three sightlines probed the `super-virial' gas in the Northern hemisphere. \citealt{Lara-DI2023,Lara-DI2024a} stacked $47$ sightlines across the sky and detected this super-virial phase using \sixiv and \sxvi absorption lines. In this work, we have extended the search for the hot gas in X-ray absorption towards another sightline with high quality grating spectra, PKS\,2155-304 that lies in the Southern latitudes. 

PKS\,2155-304 is well observed by \emph{Chandra} (\citealt{Weisskopf2000}) and \emph{XMM-Newton} (\citealt{Jansen2001}) for calibration purposes and studying the blazar itself. The deep archival data obtained from these sensitive instruments across multiple sightlines offer a valuable opportunity to study the Milky Way CGM using X-ray absorption line spectroscopy (\citealt{Mathur2022}), allowing tighter constraints on its physical properties. We utilize archival data for our study and searched for absorption lines of Hydrogen and Helium-like ions of Carbon, Nitrogen, Oxygen, Neon, Magnesium, and Silicon.

The paper is organized as follows. In Section \ref{sec:data_red_and_an}, we discuss our data reduction and analysis method, followed by results in Section \ref{sec:results}. Finally, we discuss and conclude our work in Sections \ref{sec:discussion} and \ref{sec:conclusion}.

\begin{table*}
    \centering
    \setlength{\tabcolsep}{4pt}
  \caption{Detected/non-detected ions and their EW for various instruments.}
    \begin{tabular}{|c|c|c|c|c|c|c|c|}
    \hline
        \multirow{2}{3em}{Ion} & \multirow{2}{5em}{Transition} & \multirow{2}{3em}{$\lambda_0$ ({\r A})} & \multicolumn{5}{c|}{EW (m{\r A})}  \\ 
        \cline{4-8}
        & & & HRC-LETG & ACIS-LETG & ACIS-MEG & RGS1 & RGS2 \\
        (1) & (2) & (3) & (4) & (5) & (6) & (7) & (8)  \\
        \hline
        
        \cv & K$\alpha$ & $40.267$ & $7.3\pm 3.3$ & \hbox{--} & \hbox{--} & \hbox{--} & \hbox{--} \\
        
        \cvi & K$\alpha$ & $33.736$ & $< 9.8$ & $6.3\pm 3.1$ & \hbox{--} & $6.2^{+2.0}_{-1.7}$ & $5.9\pm 1.6$ \\
        
        \nvi & K$\alpha$ & $28.787$ & $5.2\pm 2.0$ & $< 10.5$ & \hbox{--} & cool pixel & $< 6.6$ \\
       
        \nvii & K$\alpha$ & $24.781$ & $< 5.3$ & $3.2\pm 1.4$ & \hbox{--} & chip gap & $< 4.3$ \\ 

        \oiv & K$\alpha$ & $22.741$ & $5.8^{+2.1}_{-2.0}$ & $4.9\pm 1.6$ & \hbox{--} & node gap & dead CCD \\
        
        \ov & K$\alpha$ & $22.370$ & $<9.0$ & $<4.2$ & $<15.4$ & $<5.6$ & dead CCD \\
        
        \ovii & K$\alpha$ & $21.601$ & $10.6\pm 1.9$ & $9.5^{+1.0}_{-1.9}$ & $13.5^{+3.4}_{-2.1}$ & $15.4^{+1.2}_{-1.1}$ & dead CCD \\
        
         & K$\beta$ & $18.627$ & $< 7.5$ & $2.1^{+1.2}_{-1.0}$ & $< 9.0$ & $4.7^{+1.0}_{-1.3}$ & $<1.8$ \\
         
        \oviii & K$\alpha$ & $18.969$ & $5.1^{+1.4}_{-1.8}$ & $3.6\pm 1.1$ & $6.6\pm 2.0$ & cool pixel & $7.6\pm 1.1$ \\
        
        \neix & K$\alpha$ & $13.447$ & $6.6^{+1.7}_{-1.6}$ & $< 4.0$ & $3.4^{+1.1}_{-1.3}$ & dead CCD & $7.0^{+1.2}_{-1.1}$ \\
        
        \nex & K$\alpha$ & $12.134$ & $< 7.9$ & $< 3.3$ & $< 3.4$ & dead CCD & $< 4.0$ \\
       
        \mgxi & K$\alpha$ & $9.169$ & $< 7.2$ & $< 1.6$ & $< 0.7$ & $< 4.3$ & $5.7\pm 1.7$ \\
       
        \mgxii & K$\alpha$ & $8.421$ & $< 8.2$ & $< 2.6$ & $1.5\pm 0.6$ & $< 5.9$ & $< 9.1$ \\
      
        \sixiii & K$\alpha$ & $6.648$ & $< 3.1$ & $< 2.4$ & $< 2.5$ & $< 6.3$ & $<8.5$  \\
        
        \sixiv & K$\alpha$ & $6.182$ & $3.7 \pm 1.8$ & $< 2.5$ & $< 2.6$ & $< 18.8$ & $<24.9$  \\
        
        \hline
    \end{tabular}
     \label{tab:all_instruments}
\end{table*}

\section{Data Reduction and Analysis}
\label{sec:data_red_and_an}

PKS\,2155-304 is a well-studied Blazar ($l=17.7^{\circ}$, $b=-52.2^{\circ}$) at $z=0.116$ (\citealt{Falomo1993}). There are several archival \emph{Chandra}\footnote{\url{https://cda.harvard.edu/chaser/}} and \emph{XMM-Newton}\footnote{\url{https://heasarc.gsfc.nasa.gov/docs/archive.html}} observations available towards this sightline. We use the grating data of these observatories to look for $z=0$ absorption lines of H-like and He-like metal ions. We analyze ACIS-MEG, ACIS-LETG, and HRC-LETG data of \emph{Chandra} and RGS1 and RGS2 data of \emph{XMM-Newton}.   

\subsection{Data Reduction}\label{sec:data_red}

We extract all the archival data of ACIS-LETG\footnote{ObsID: 1015, 1703, 1790, 1795, 1796, 1797, 2323, 2324, 2335, 3168, 3667, 3668, 3669, 3707, 4416, 6090, 6091, 6874, 6924, 6927, 7293, 8388, 9704, 9706, 9708, 9710, 9713, 10662, 11965, 13096, 15475, and 16423.}, HRC-LETG\footnote{ObsID: 331, 1013, 1704, 3166, 3709, 4406, 5172, 6922, 6923, 7294, 7295, 8379, 9707, 9709, and 9711.}, ACIS-MEG\footnote{ObsID: 337, 1014, 1705, 3167, 3706, 3708, 5173, 6926, 7291, 8380, 8436, 9705, and 9712.} and RGS1/RGS2\footnote{ObsID: 0080940101, 0080940301, 0080940401, 0124930101, 0124930201, 0124930301, 0124930501, 0124930601, 0158960101, 0158960901, 0158961001, 0158961101, 0158961301, 0158961401, 0411780101, 0411780201, 0411780301, 0411780401, 0411780501, 0411780601, 0411780701, 0411780801, 0411781301, 0411781701, 0411782101, 0727770101, 0727770501, 0727770901, and 0727771301.} for our analysis. 

For \emph{Chandra} data, we process and reduce the data for each observation with the latest calibration database of \texttt{CIAO} v4.16 using the \texttt{Chandra\_repro} command. We combine the positive $1^{\rm st}$ order spectra for all the observations and similarly, negative $1^{\rm st}$ order spectra for all the observations. We then combine the co-added positive and negative order spectra individually for ACIS-LETG, HRC-LETG, and ACIS-MEG using the \texttt{combine\_grating\_spectra} command.
In HRC-LETG, the spectrum is the combination of all the orders ($1^{\rm st}$ to $8^{\rm th}$), as HRC cannot resolve the individual diffracted orders. To account for the contribution of all the orders, we add the positive and negative order RMF and ARF files of all the observations for each order ($1$ to $8$) using \texttt{addresp} command. We load the response files corresponding to all the orders for our HRC-LETG analysis.

For \emph{XMM-Newton} RGS1/RGS2 data, we use the latest calibration data \texttt{CALDB} v4.11.2 and reduce the data using the \texttt{rgsproc} command. We then determine the Good Time Interval (GTI) using the \texttt{tabgtigen} command by removing time intervals where the background count rate exceeds $0.1 \, \rm cnt \, s^{-1}$. After obtaining GTI, we combine the first-order spectra of all the observations using the \texttt{rgscombine} command separately for RGS1 and RGS2. We only consider the $1^{\rm st}$- order spectra for all the instruments (except HRC-LETG), as the effective area is lower for higher-order spectra.

\begin{table*}
    \centering
    \setlength{\tabcolsep}{3pt}
   \renewcommand{\arraystretch}{1.2}
   \caption{Parameters of the absorption lines used for our analysis. Fourth column: detected EW; fifth, sixth, and seventh columns: column density contribution from the three distinct temperature phases calculated using hybrid-ionization modeling; eighth column: total column density across three phases; ninth column: column density from detected EW using the linear part of the curve of growth.}
    \begin{tabular}{|c|c|c|c|c|c|c|c|c|}
    \hline
     (1) & (2) & (3) & (4) & (5) & (6) & (7) & (8) & (9) \\
     Ion & Transition & Instrument & EW & $N_{T_1}$ & $N_{T_2}$ & $N_{T_3}$ & $N_{\rm tot}$ & $N_{\rm EW}$ \\
      &  & &  (m{\r A}) & (cm$^{-2}$) & (cm$^{-2}$) & (cm$^{-2}$) & (cm$^{-2}$) & ($\times 10^{15}$ cm$^{-2}$) \\ 
     
      \hline
     \cv  & K$\alpha$ & HRC-LETG & $7.27\pm 3.31$ & $1.3\pm 0.5 \times 10^{15}$ & $9.8^{+12.2}_{-1.7}\times 10^{13}$ & \hbox{--} & $1.4^{+0.62}_{-0.52}\times 10^{15}$ & $0.78\pm 0.36$ \\
         
      \cvi & K$\alpha$ & RGS2 & $5.94 \pm 1.60$ & $4.3^{+0.8}_{-1.2} \times 10^{12}$ & $8.3^{+7.1}_{-2.6}\times 10^{14}$ & $4.4^{+3.6}_{-1.0}\times 10^{14}$ & $1.3^{+1.1}_{-0.4}\times 10^{15}$ & $1.42 \pm 0.38$ \\
      
      \nvi & K$\alpha$ & HRC-LETG & $5.18\pm 1.96$ & $2.7^{+2.7}_{-0.5} \times 10^{14}$ & $3.8^{+3.8}_{-1.0}\times10^{14}$ & \hbox{--} & $6.5^{+6.5}_{-1.5}\times10^{14}$ & $1.05\pm 0.40$ \\
         
      \nvii & K$\alpha$ & HRC-LETG & $< 5.26$ & $3.2^{+4.2}_{-0.2} \times 10^{11}$ & $7.7^{+3.1}_{-0.4}\times 10^{14}$ & $3.8^{+3.1}_{-0.9}\times 10^{14}$ & $1.2^{+0.6}_{-0.1}\times 10^{15}$ & $< 2.33$ \\ 
      
      \ovii & & & & $5.0^{+16.9}_{-4.9} \times 10^{13}$ & $4.4^{+2.2}_{-0.5}\times 10^{15}$ & \hbox{--} & $4.5^{+2.4}_{-0.5}\times 10^{15}$ & \sout{$9.32$}$^{+1.81}_{-0.76}$\footnote{Saturation corrected column density. It cannot be used in analysis, as \ovii has a significant contribution from various temperature phases.} \\
       & K$\alpha$ & ACIS-LETG & $9.50^{+0.96}_{-1.88}$ &  & & &  & $3.31^{+0.34}_{-0.65}$ \\
         & K$\beta$ & ACIS-LETG & $2.06^{+1.22}_{-1.00}$ &  & & &  & $4.59^{+2.72}_{-2.22}$ \\
         
      \oviii & K$\alpha$ & ACIS-LETG & $3.60\pm 1.11$ & \hbox{--} & $1.9^{+1.3}_{-0.4}\times 10^{15}$ & $3.1^{+2.8}_{-0.7}\times 10^{15}$ & $5.0^{+4.1}_{-1.1}\times 10^{15}$ & $2.72\pm 0.84$ \\
      
      \neix & K$\alpha$ & HRC-LETG & $6.62^{+1.67}_{-1.64}$ & -- & $1.1^{+0.2}_{-0.6}\times 10^{16}$ & \hbox{--} & $1.1^{+0.2}_{-0.6}\times 10^{16}$ & $5.71^{+1.44}_{-1.42}$ \\
         
      \nex & K$\alpha$ & HRC-LETG & $<7.91$ & -- & $1.3\pm 0.7 \times 10^{14}$ & $5.2^{+5.9}_{-0.7}\times 10^{15}$ & $5.3^{+6.0}_{-0.8} \times 10^{15}$ & $<14.59$ \\ 
      
      \mgxi & K$\alpha$ & ACIS-MEG & $<0.66$ & -- & $4.7^{+0.9}_{-2.5}\times 10^{14}$ & \hbox{--} & $4.7^{+0.9}_{-2.5}\times 10^{14}$ & $<1.19$ \\
      
      \mgxii & K$\alpha$ & ACIS-MEG & $1.47\pm 0.59$ & -- & \hbox{--} & $6.1^{+8.0}_{-0.5}\times 10^{15}$ & $6.1^{+8.0}_{-0.5}\times 10^{15}$ & $5.44\pm 2.27$ \\
      
      \sixiii & K$\alpha$ & HRC-LETG & $<3.07$ & -- & $4.69^{+5.05}_{-1.11}\times 10^{14}$ & \hbox{--} & $4.69^{+0.99}_{-0.26}\times 10^{14}$ & $<10.35$ \\
      
      \sixiv & K$\alpha$ & HRC-LETG & $3.73^{+1.75}_{-1.78}$ & -- & \hbox{--} & $5.5^{+7.0}_{-0.5}\times 10^{16}$ & $5.5^{+7.0}_{-0.5}\times 10^{16}$ & $26.54^{+12.44}_{-12.63}$ \\
      \hline
    
    \end{tabular}
    
    \label{tab:detection}
\end{table*}

After obtaining all the spectra ($3$ from \emph{Chandra} and $2$ from \emph{XMM-Newton}), background spectra, response matrix file (RMF), and ancillary response file (ARF), we load them in \texttt{XSPEC} v12.13.0 for further analysis. The total exposure times for ACIS-LETG, HRC-LETG, ACIS-MEG, RGS1/RGS2 are $718.1, 319.5, 289.4$, and $1981$ Ks, respectively. We use $\chi^2$ statistics for our analysis throughout the work. We quote $1\sigma$ error for detections and $3\sigma$ upper limit for non-detections.

\begin{figure*}
    \centering
    \includegraphics[width=0.3\textwidth]{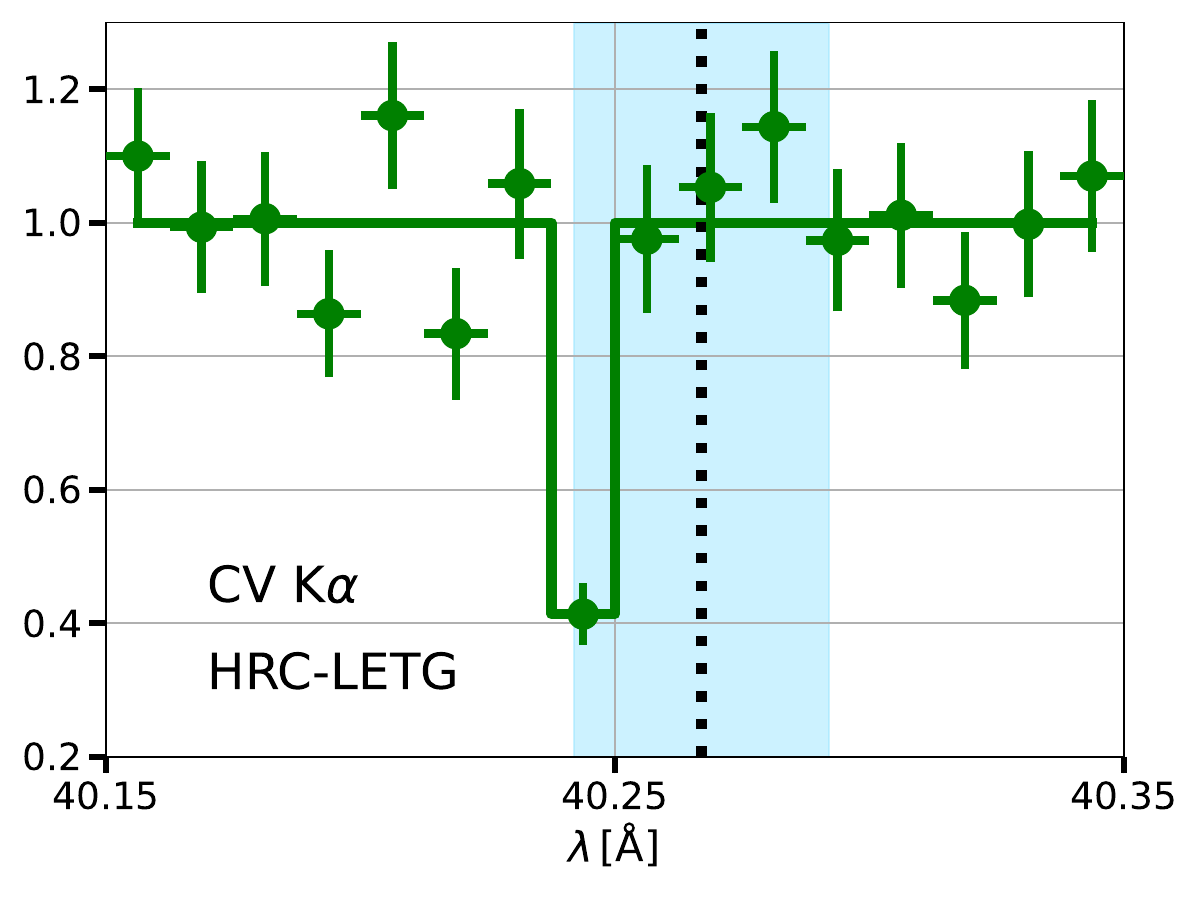}
    \includegraphics[width=0.3\textwidth]{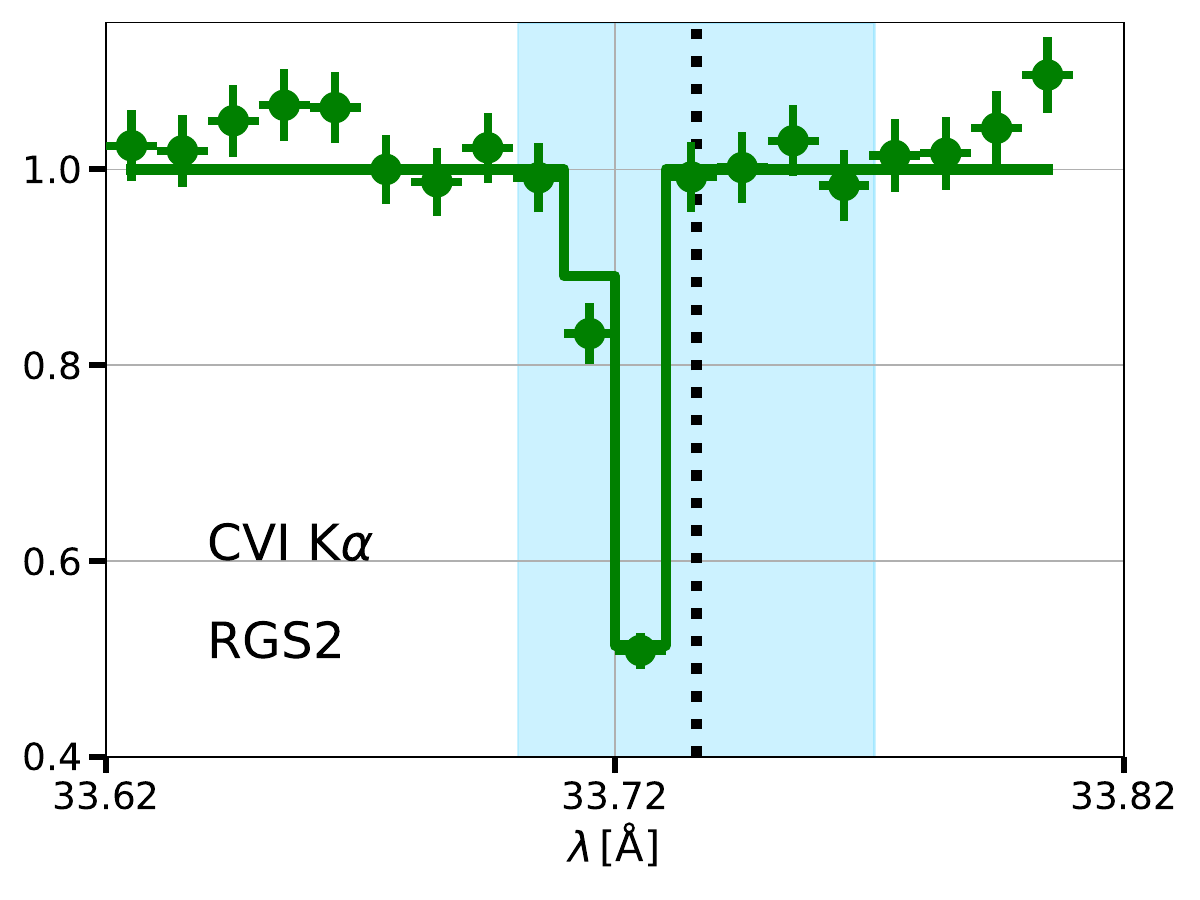}
    \includegraphics[width=0.3\textwidth]{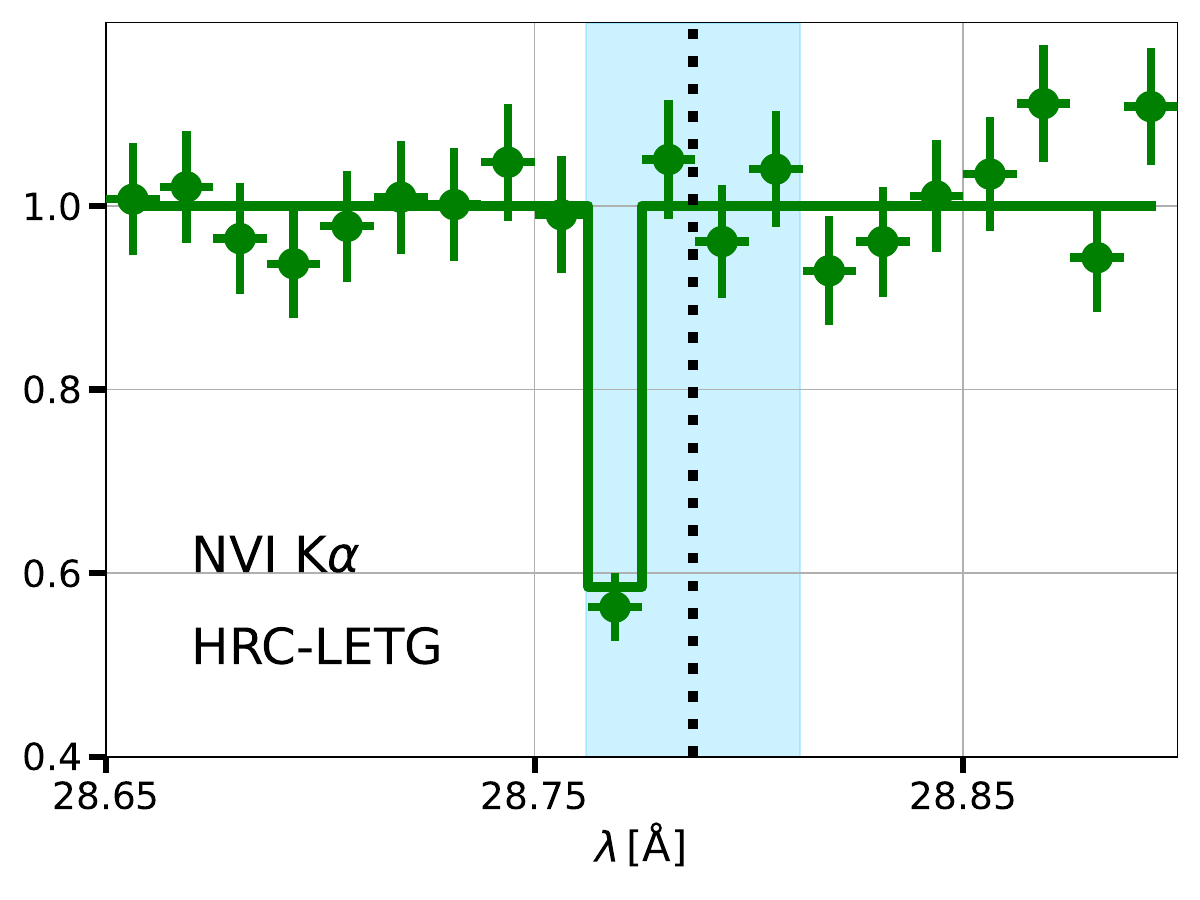}
    \includegraphics[width=0.3\textwidth]{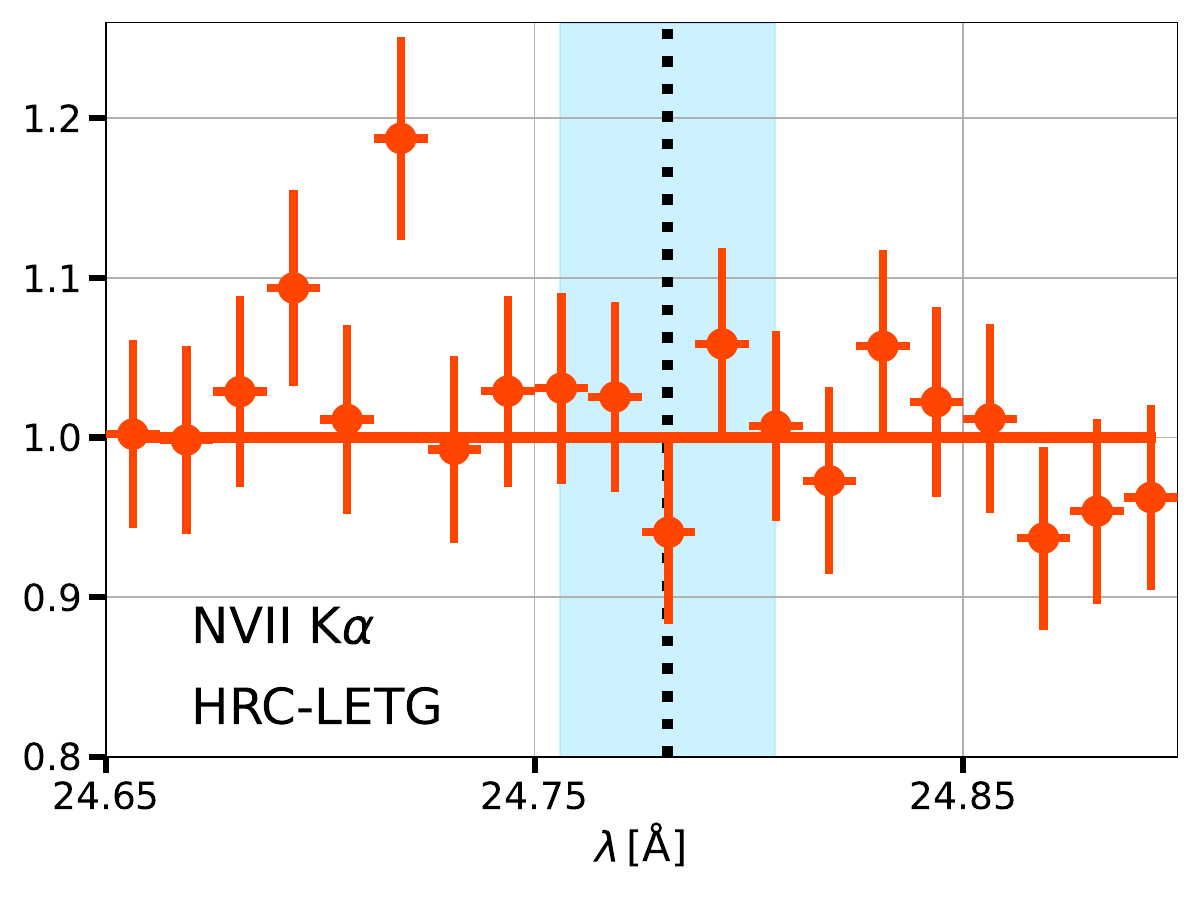}
    \includegraphics[width=0.3\textwidth]{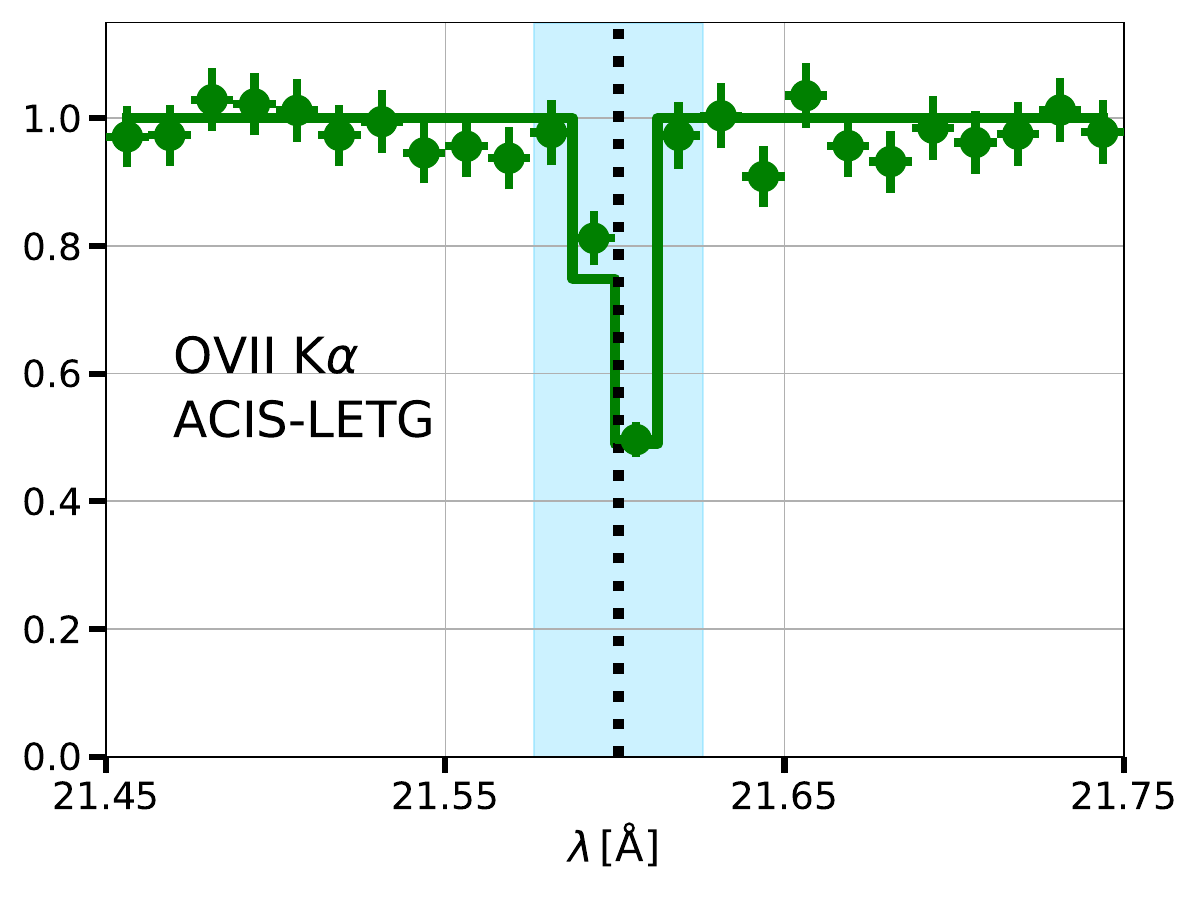}
    \includegraphics[width=0.3\textwidth]{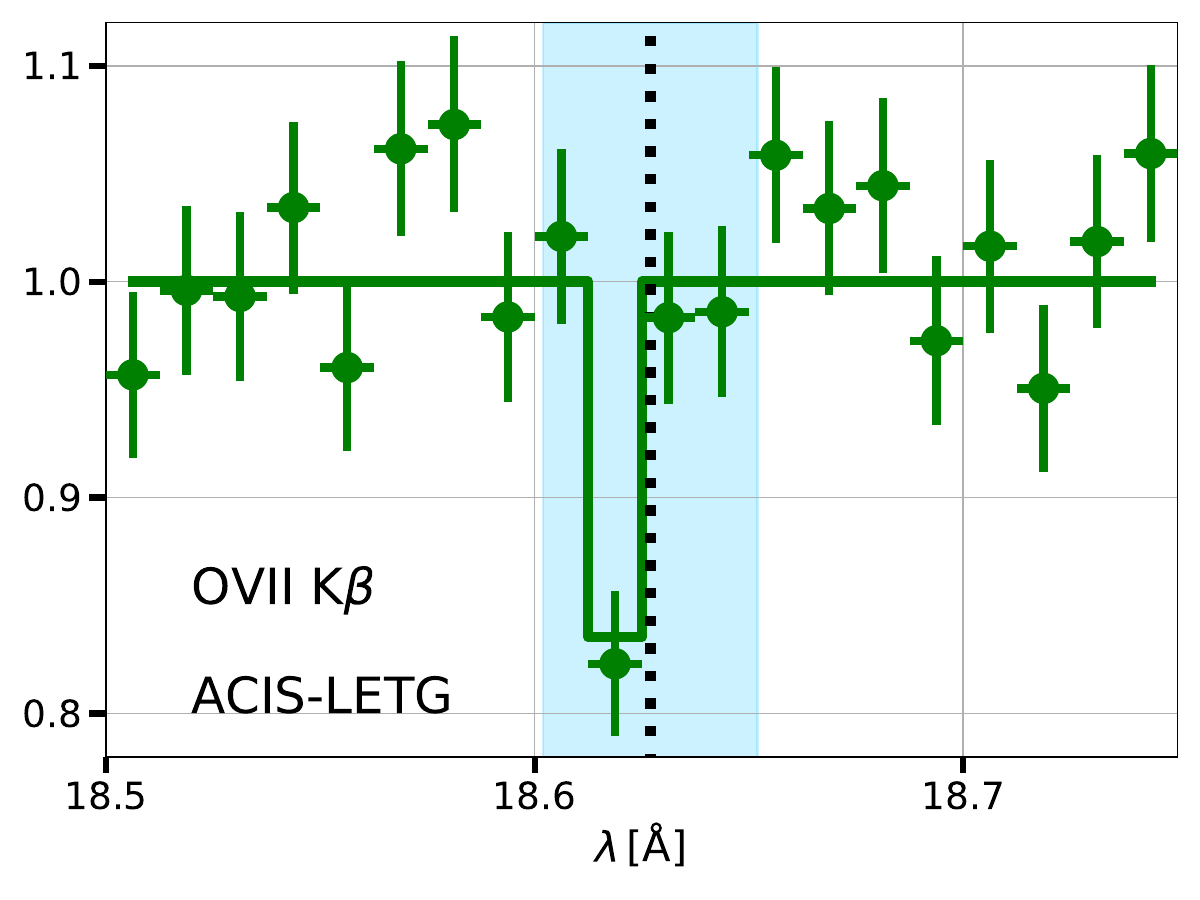}
    \includegraphics[width=0.3\textwidth]{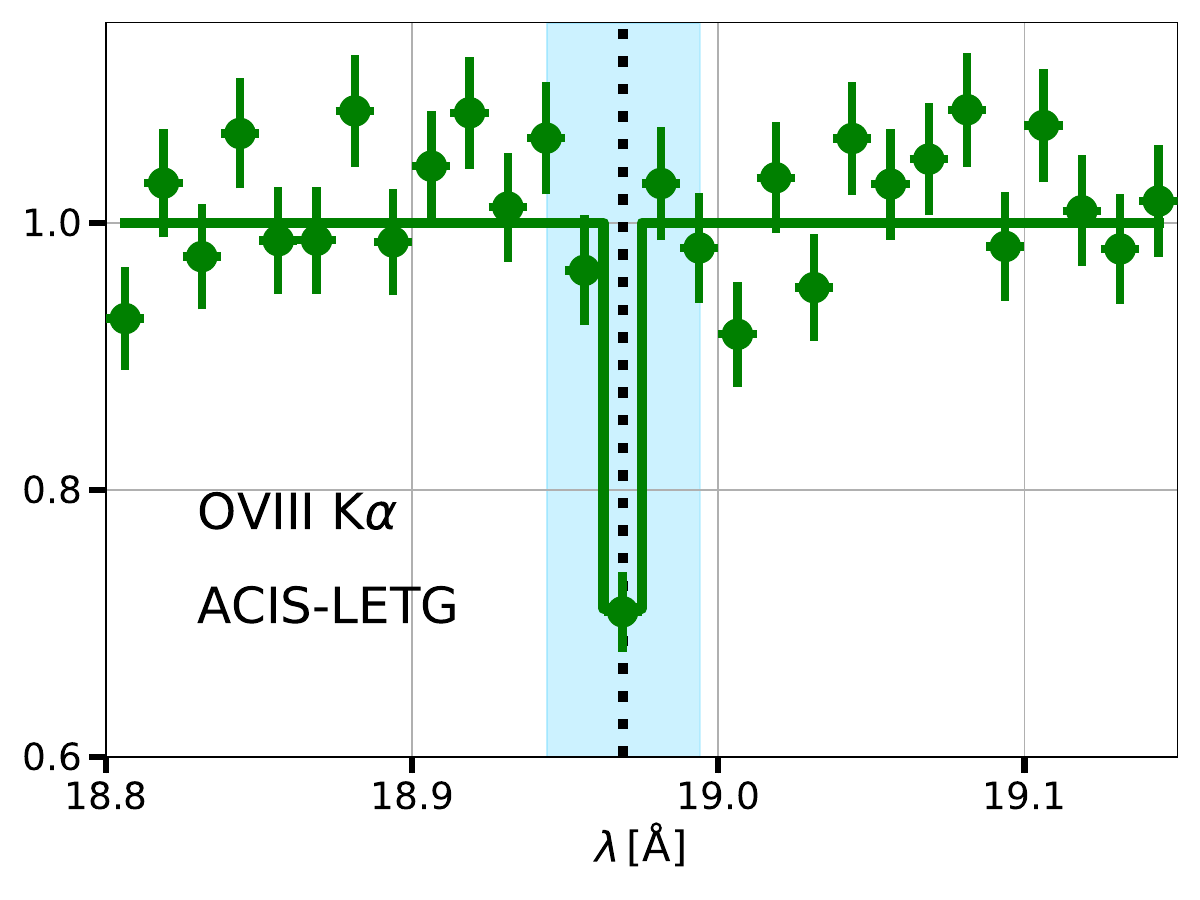}
    \includegraphics[width=0.3\textwidth]{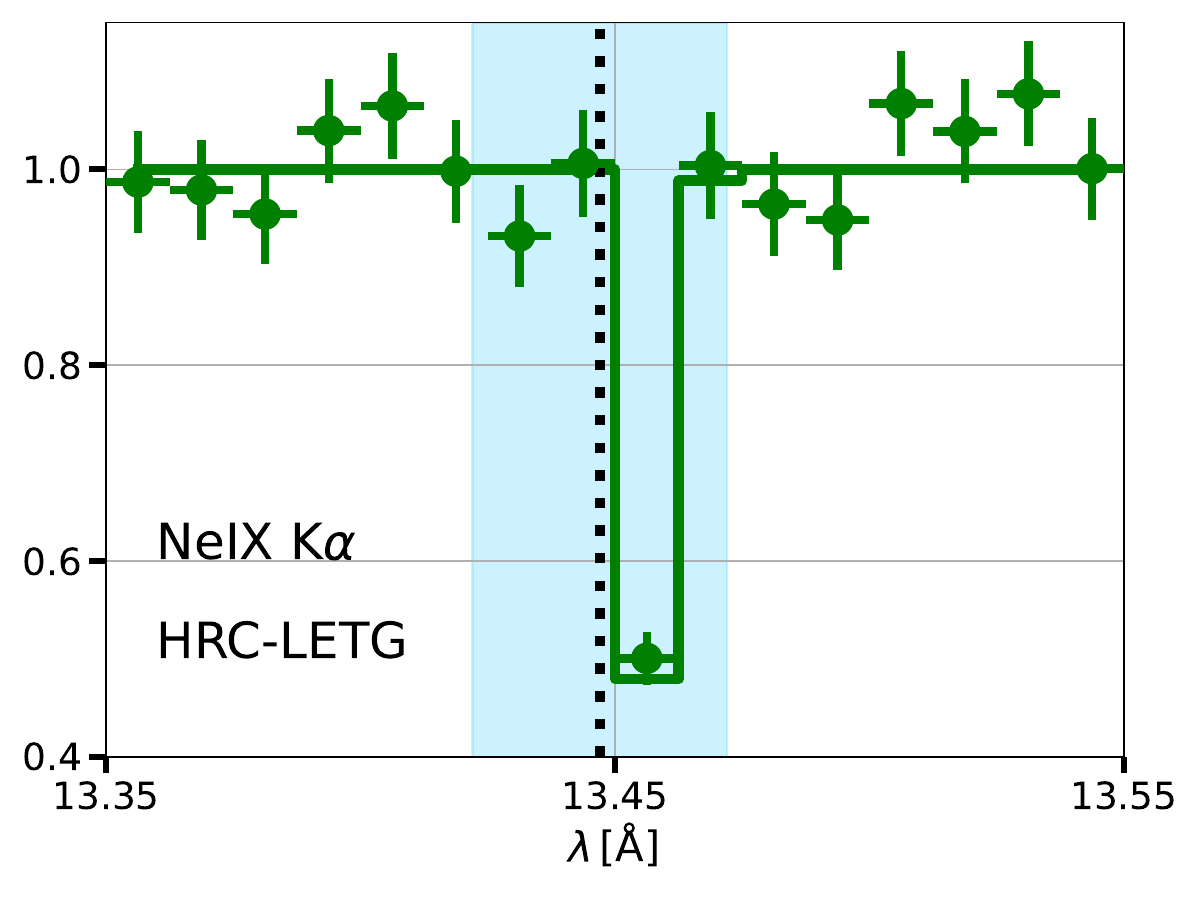}
    \includegraphics[width=0.3\textwidth]{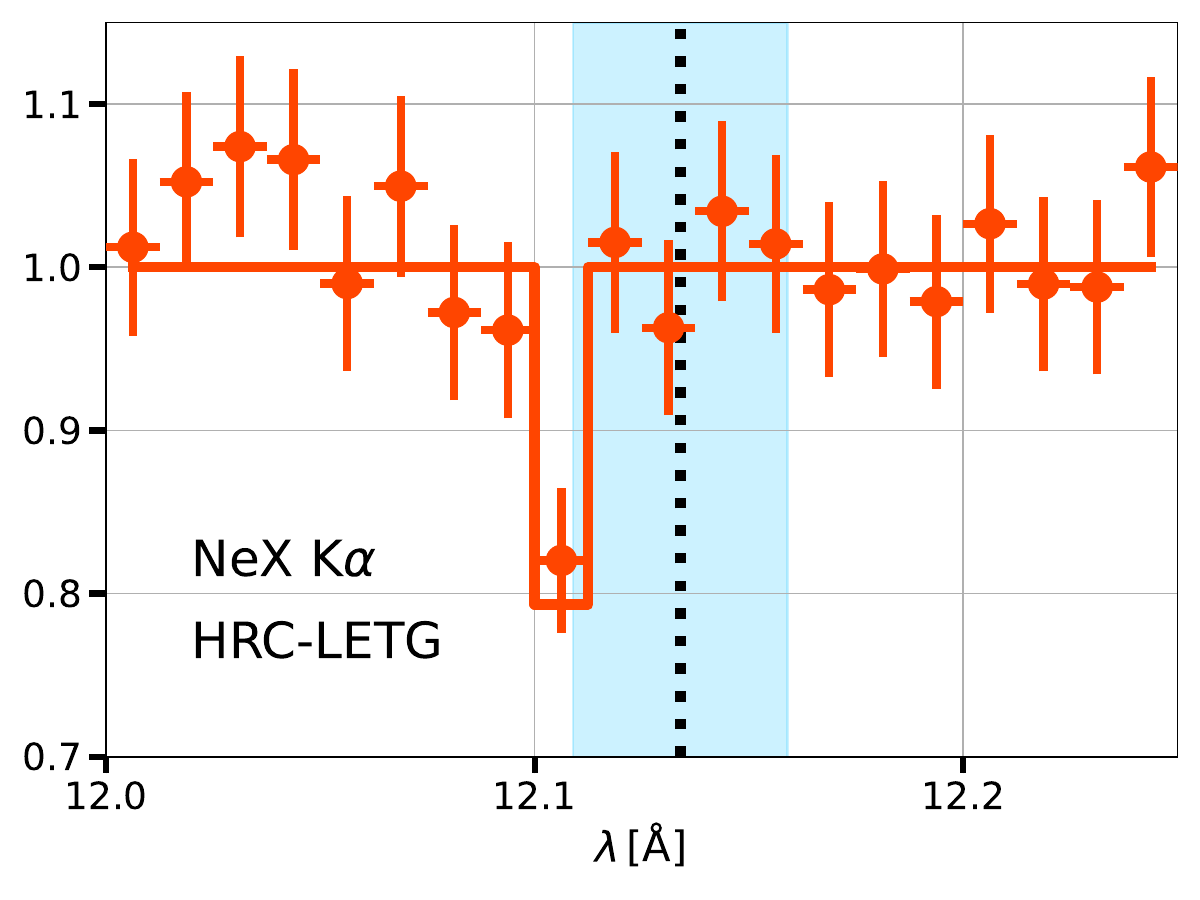}
    \includegraphics[width=0.3\textwidth]{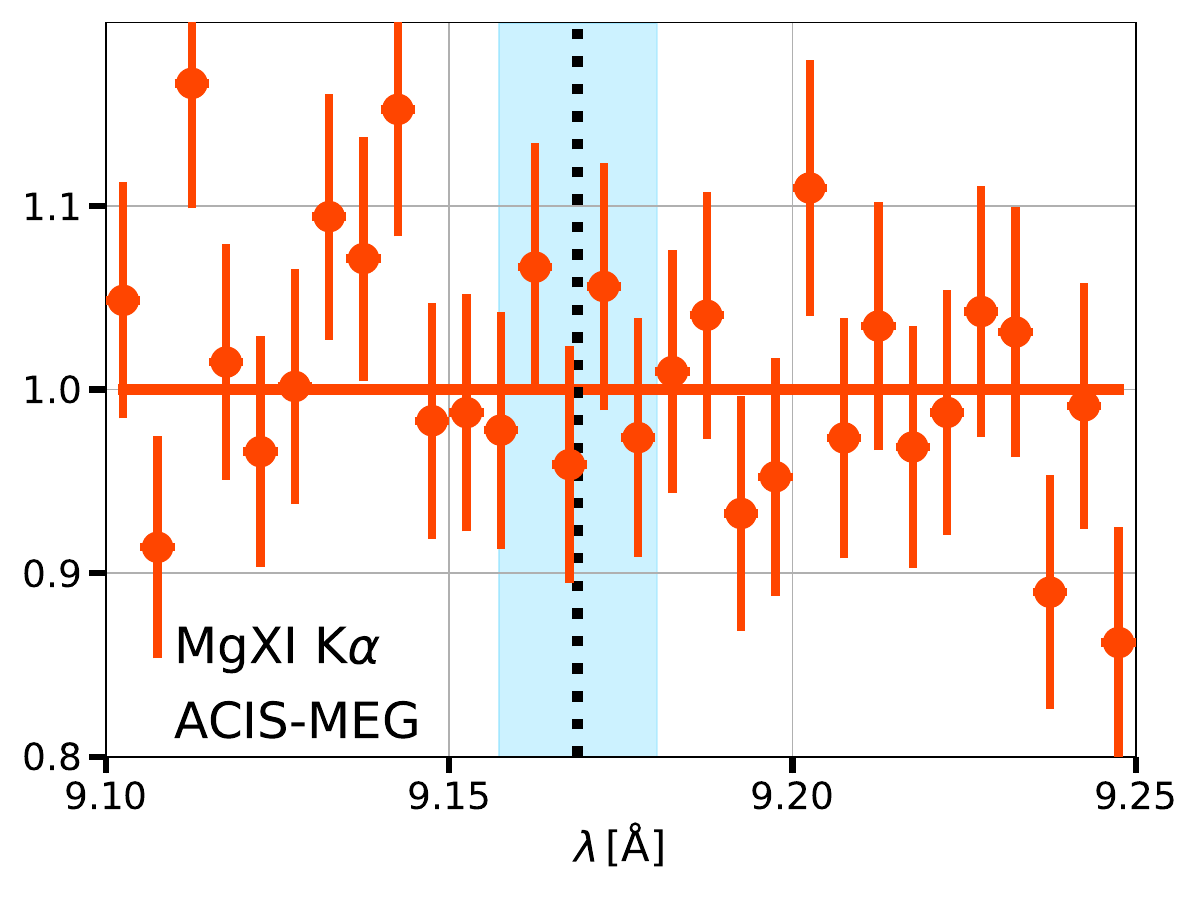}
    \includegraphics[width=0.3\textwidth]{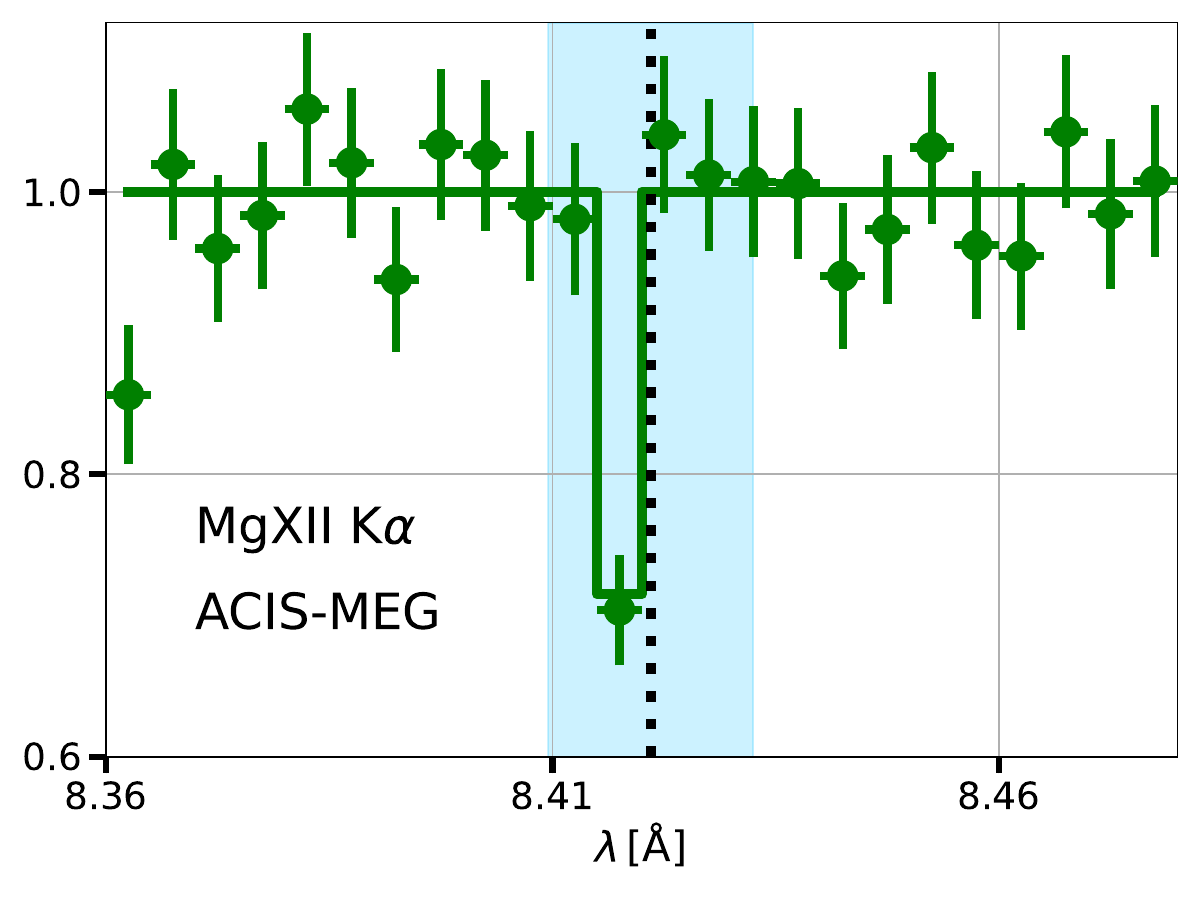}
    \includegraphics[width=0.3\textwidth]{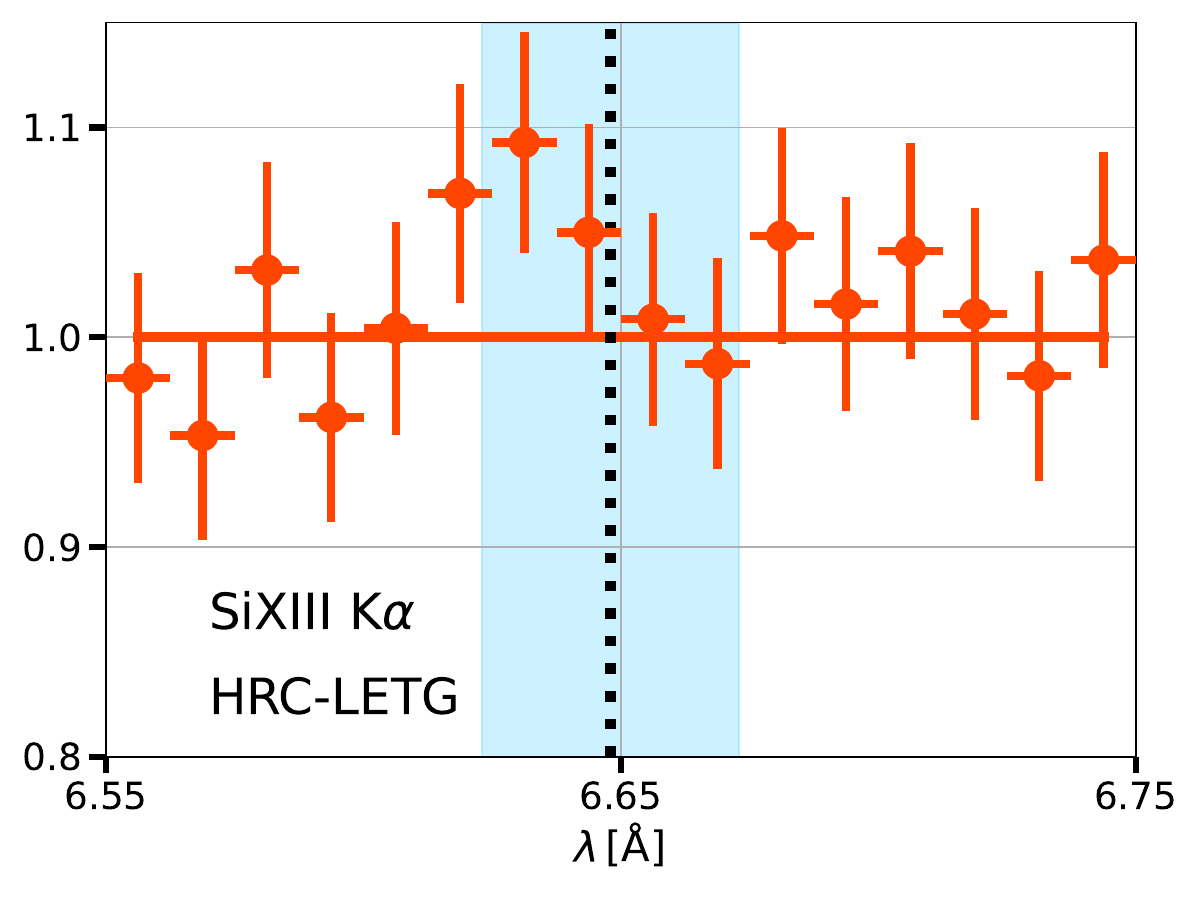}
    \includegraphics[width=0.3\textwidth]{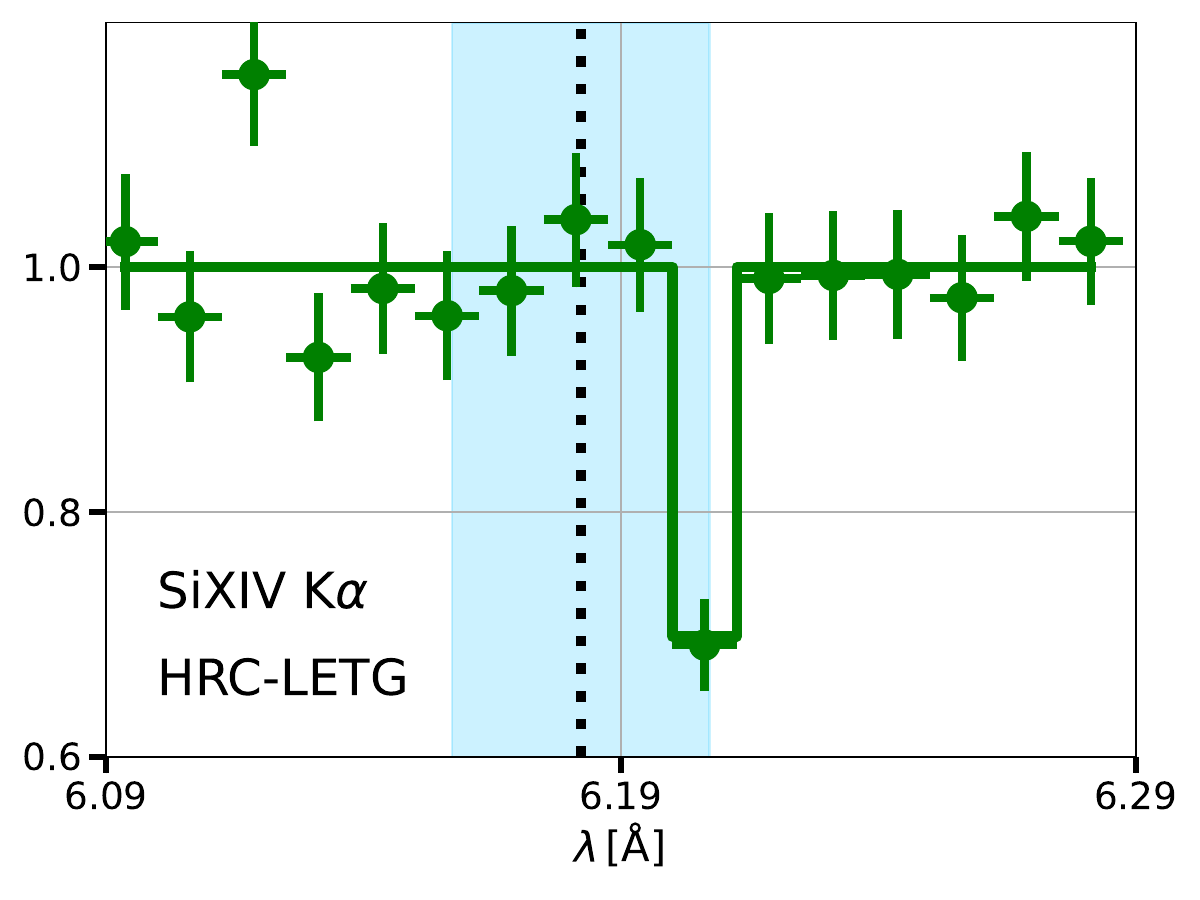}
    \includegraphics[width=0.3\textwidth]{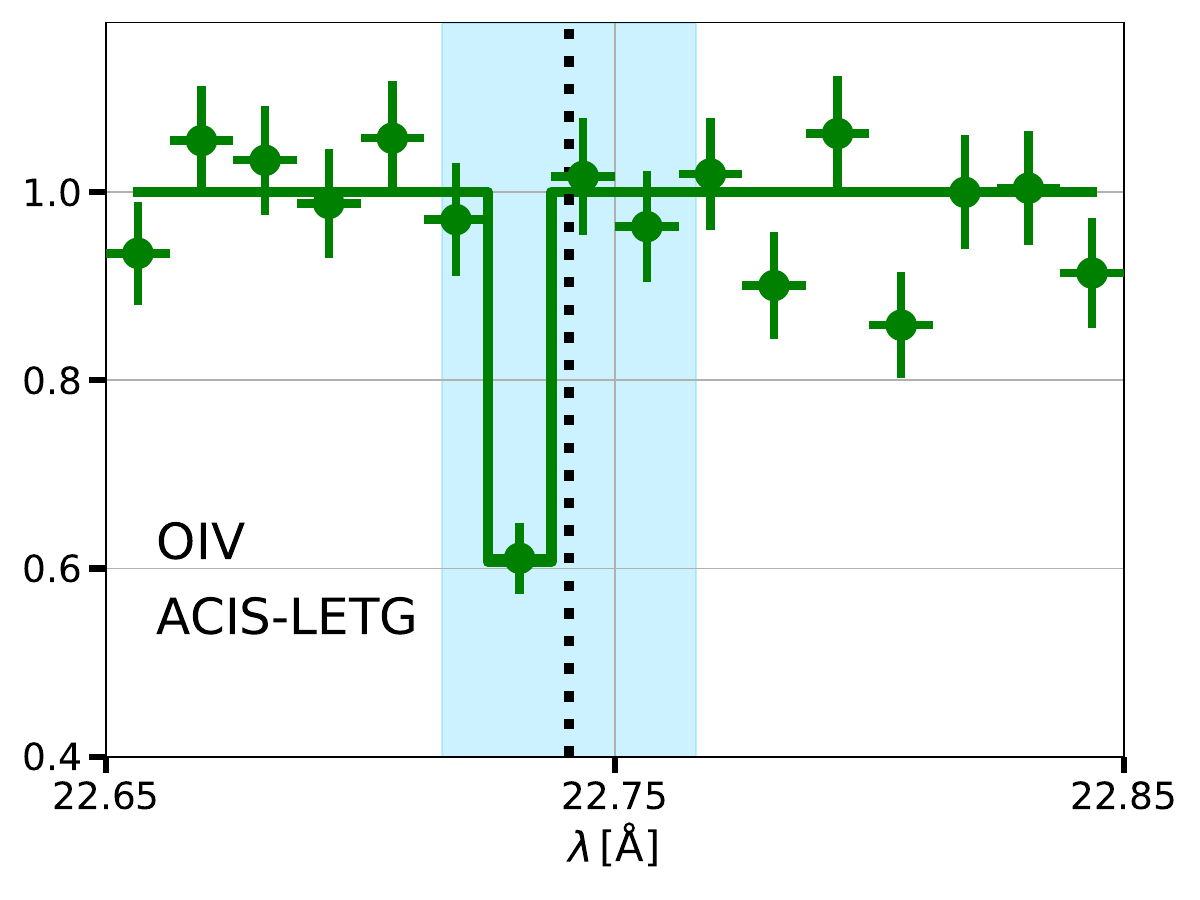}
    \includegraphics[width=0.3\textwidth]{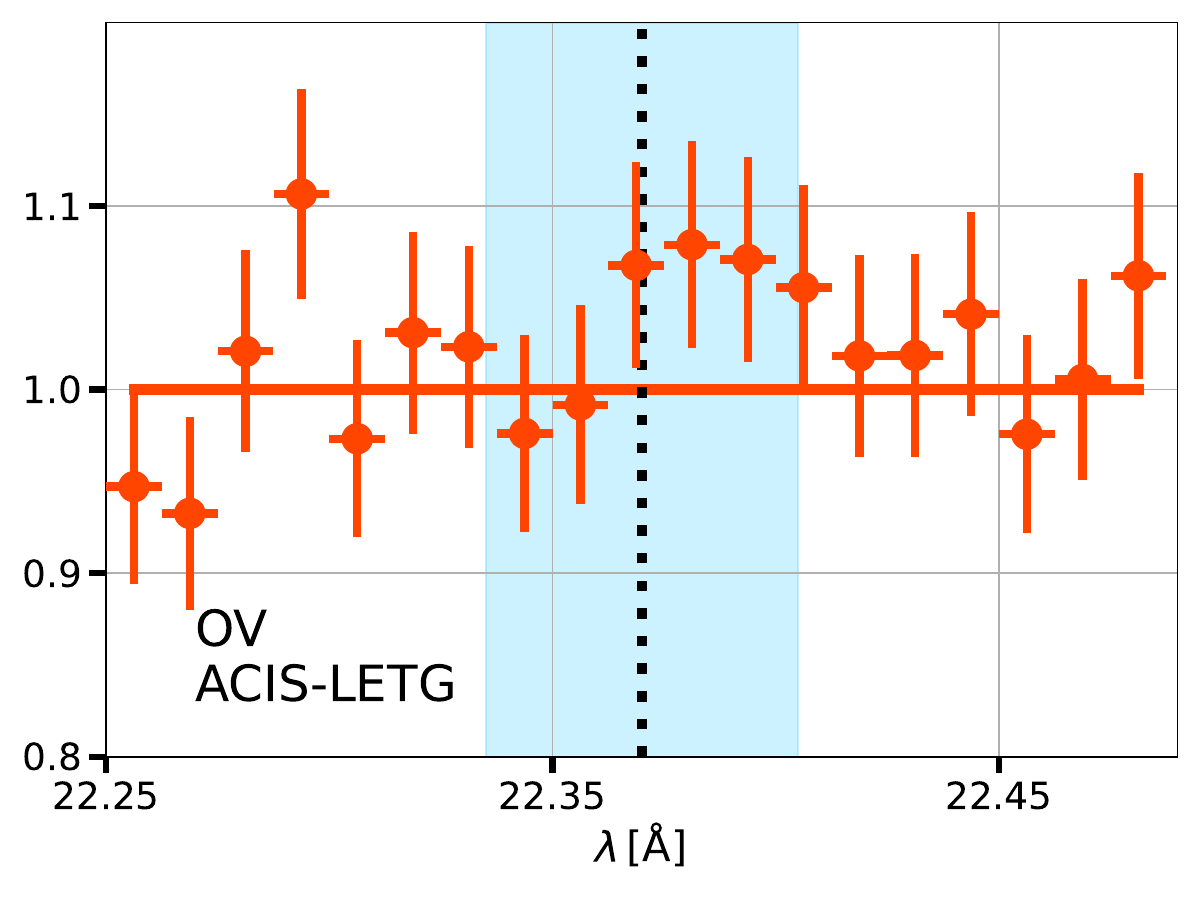}
    \caption{Normalized absorption profile of the detected (green) and non-detected (red) ions. The dotted black vertical line in each panel shows the $z=0$ value of the line center of the respective ions. The shaded blue region in each panel indicates the resolution element of the respective instrument relative to the $z=0$ value of the line center of the ion.}
    \label{fig:detection}
\end{figure*}

\subsection{Data Analysis}
\label{sec:model_ind_res}

After obtaining the spectra, we model the continuum with a power-law absorbed by our Galactic disk using Tuebingen-Boulder ISM absorption model \texttt{tbabs$\times$powerlaw}. \texttt{tbabs} calculates the cross section for X-ray absorption by the ISM as the sum of the cross sections for X-ray absorption due to the gas-phase ISM, the grain-phase ISM, and the molecules in the ISM. We fix the equivalent Hydrogen column density for \texttt{tbabs} model to $N_{\rm H}=1.3\times 10^{20} \, \rm cm^{-2}$ \citep{HI4PICollaboration2016}. We let the index and normalization of the power-law vary. Next, we search for the H/He-like ions of Carbon, Nitrogen, Oxygen, Neon, Magnesium, and Silicon. To do so, we select $\approx \pm 2$ {\r A} wavelength range around the $z=0$ values of strong lines of these metals. To account for the metal absorption line, we add an absorbing Gaussian model (\texttt{agauss}). Therefore, our overall model is \texttt{tbabs*(po+agauss)}. For the line center of the Gaussian, we start with the $z=0$ value of the absorption line and allow it to vary within the resolution element\footnote{Resolution element ($\Delta \lambda$, FWHM) of $23$, $50$, and $70$ m{\r A} for MEG, LETG, and RGS, respectively.} of the respective grating. We fix the width of the Gaussian to $10^{-5}$ {\r A} as the absorption lines are not expected to be resolved. We allow the normalization of the Gaussian to vary.
We calculate the equivalent width (EW) of the absorption line using the \texttt{eqw} command. To calculate the $1\sigma$ uncertainty in EW, we first calculate $1\sigma$ error in normalization of the Gaussian using the \texttt{error} command. We then fix the normalization of the Gaussian to the $1\sigma$ upper and lower values and calculate the EW corresponding to these values, which is the $1\sigma$ uncertainty in EW. We call it a detection if the significance of EW is more than $2\sigma$,\footnote{Significance is defined as the ratio of best-fit EW to the $1\sigma$ error on the lower side}; otherwise, we quote a $3\sigma$ upper limit on the EW.
We repeat the above analysis for strong lines of each ion and each instrument and list our results in Table \ref{tab:all_instruments}. 

We show the EW of detected and non-detected ions in Table \ref{tab:all_instruments} for all the instruments. 
We detect (significance $\geq 2\sigma$) several absorption lines in different instruments. There is no wavelength coverage in MEG for C and N lines.  \cv K$\alpha$ is detected only in HRC-LETG, as there is no wavelength coverage in other instruments. 
\cvi K$\alpha$ detection in ACIS-LETG, RGS1, and RGS2 are consistent with each other within $1\sigma$ uncertainty, while the $3\sigma$ upper limit from HRC-LETG is also consistent with different instruments.
\nvi K$\alpha$ is detected only in HRC-LETG, while ACIS-LETG and RGS2 show non-detection. For RGS1, there is a cool pixel at the corresponding wavelength. Detection and non-detections are consistent with each other.
We detect \nvii K$\alpha$ only in ACIS-LETG, while HRC-LETG and RGS2 show non-detection, and in RGS1, there is a chip gap at this wavelength. Again, detection and non-detections are consistent across instruments.
\oiv K$\alpha$ is detected in HRC-LETG and ACIS-LETG, and they are consistent with each other. We do not detect \ov in any of the instruments.
\ovii K$\alpha$ is detected in all the instruments except for RGS2 due to a dead CCD ($\approx 20\hbox{--}24$ {\r A}). The EW value in ACIS-LETG is the lowest among all the instruments and is consistent only with HRC-LETG within $1\sigma$ uncertainty. The EW values in RGS1 and MEG are high and are consistent with each other but not with other instruments together.
\ovii K$\beta$ is detected in ACIS-LETG and RGS1, while there is no detection in other instruments, which are consistent with each other within error bars except for RGS1.
\oviii K$\alpha$ shows a similar trend as \ovii K$\alpha$, whereas there is cool pixel at this wavelength for RGS1. 

\neix K$\alpha$ is detected in HRC-LETG, MEG, and RGS2, while ACIS-LETG shows non-detection. The values are not consistent across instruments. Neon lines cannot be detected in RGS1 due to a dead CCD at $\approx 10\hbox{--}14$ {\r A}.
\nex K$\alpha$ is not detected in any of the instruments.
\mgxi K$\alpha$ is not detected in any of the instruments except in RGS2. MEG, which is most sensitive at these wavelengths, shows a non-detection with a $3\sigma$ upper limit of $0.7$ m{\r A}, while the EW obtained using RGS2 is very high. This could be due to the effect of a nearby cool pixel.
\mgxii K$\alpha$  is detected in the CGM of the MW for the first time. It is detected in MEG, and the upper limits from other instruments are consistent with detection.
\sixiii K$\alpha$ is not detected in any instruments. However, \sixiv K$\alpha$ is detected in HRC-LETG, with non-detection in other instruments, which are consistent with detection within $1\sigma$ uncertainty.

Thus, we see that for most lines the measured EWs, either detections/non-detections, are consistent across instruments within $1\sigma$ uncertainties. The only exceptions are the \ovii K$\alpha$ and K$\beta$, \oviii K$\alpha$, \neix K$\alpha$, and \mgxi K$\alpha$. The inconsistencies in the EW values across instruments may be attributed to the impact of cool pixels in the surrounding wavelength regions, most notably in RGS1 and RGS2. 

\begin{figure*}
    \centering
    \includegraphics[width=\linewidth]{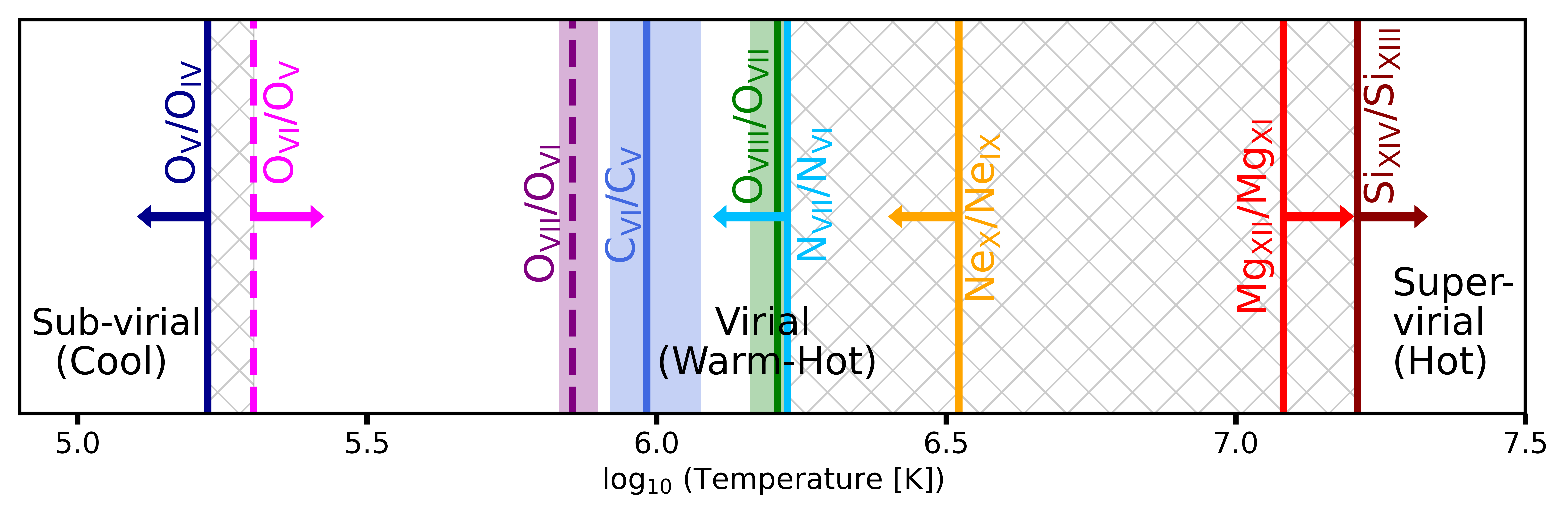}
    \caption{Temperature ranges estimated from the column density ratio of two adjacent ions of the same metal probed in X-ray absorption. It clearly shows the multiphase nature of the CGM with \textit{temperature valleys} (hatched region) in between. The dashed magenta and purple lines show the temperature estimated from the column density ratios of \ovi (detected in UV absorptions by \citealt{Collins2004}), \ov (non-detection in this work), and \ovii (this work), \ovi \citep{Collins2004} respectively.}
    \label{fig:illustration}
\end{figure*}

\subsection{Model-independent Analysis}

In Table \ref{tab:detection}, we show the detected line details in the selected instrument for each ion based on the sensitivity at the respective wavelength value and significance of detection. We use these wavelength ranges for the specific instruments in our further analysis.
The detections in the respective mentioned instruments are: \cv K$\alpha$ ($2.2\sigma$), \cvi K$\alpha$ ($3.7\sigma$), \nvi K$\alpha$ ($2.6\sigma$), \oiv K$\alpha$ ($3.1\sigma$) \ovii K$\alpha$ ($5.1\sigma$), \ovii K$\beta$ ($2.1\sigma$), \oviii K$\alpha$ ($3.2\sigma$), \neix K$\alpha$ ($4.0\sigma$), \mgxii K$\alpha$ ($2.4\sigma$), and \sixiv K$\alpha$ ($2.1\sigma$). We use the detected/non-detected ions in the mentioned instruments for our further analysis.
In Figure\,\ref{fig:detection}, we show the normalized absorption profile of the detected and non-detected ions in green and red colors, respectively. The black dotted line in each panel shows the $z=0$ wavelength value, and the blue band shows the resolution element of the instrument. The best-fit wavelength within the resolution element for all the ions is shifted from the $z=0$ value. 

By fitting the metal absorption line using the Gaussian profile, we obtain the Equivalent Width (EW), which we then convert into column density using the linear part of the curve of growth (\citealt{Draine2011}). We also detect the K$\beta$ line of \ovii along with the K$\alpha$ line. Therefore, we obtain the Doppler $b$ parameter and the saturation corrected column density for \ovii by combining the EW from K$\alpha$ and K$\beta$ lines using the flat part of the curve of growth. 

The Doppler $b$ parameter and the column density ratio of detected H to He-like ions of the same element can be used to estimate the temperature of the gas bearing these ions under Collisional Ionization Equilibrium (CIE), assuming the ions exist in the same phase. The upper limit (non-detection) on the column density also provides useful constraints on the physical conditions of the gas bearing these ions, assuming CIE.

The temperature estimated from the column density ratio of \cvi to \cv is $9.6^{+2.3}_{-1.3} \times 10^5$ K, \nvii to \nvi is $< 1.7^{+0.3}_{-0.1} \times 10^6$ K, \oiv to \ov is $<1.7 \times 10^5$ K, \ov to \ovii is $>3.3 \times 10^5$ K, \oviii to \ovii is $1.6^{+0.1}_{-0.2} \times 10^6$ K, \nex to \neix is $<3.3 \pm 0.5 \times 10^6$ K, \mgxii to \mgxi is $>1.2\pm 0.2 \times 10^7$ K, and \sixiv to \sixiii is $> 1.6^{+0.2}_{-0.3} \times 10^7$ K under CIE. These model-independent temperature estimates suggest the presence of the `super-virial' or hot gas along with expected virial or warm-hot and `sub-virial' or warm gas as shown in Figure\,\ref{fig:illustration}. A further lower temperature cool gas is also indicated by the column density ratios of \ov and \oiv. This cool phase is observed for the first time in X-ray studies. 
The dashed magenta and purple lines show the temperature estimated from the column density ratios of \ovi (detected by \citealt{Collins2004} in UV), \ov (X-ray non-detection in this work), and \ovii (X-ray detection in this work), \ovi \citep{Collins2004} respectively.
The limits on the estimated temperature also demonstrate the presence of a \textit{temperature valley} (hatched region in Figure\,\ref{fig:illustration}) between the virial gas ($\sim 10^6$ K) and the super-virial gas ($\sim 10^7$ K). A \textit{temperature valley} is also present between the sub-virial (warm) and cool phase. 

These model-independent results show the multiphase nature of the CGM with \textit{temperature valleys} in between, which roughly mimics the `multi-peaked' log-normal distribution in temperature in the CGM \citep{Dutta2024}. A `Multi-peaked' log-normal distribution is obtained by adding multiple log-normal distributions with different peaks, heights, and widths. The multiple peaks may correspond to our model-independent temperature estimates, suggesting different temperature phases in the CGM.
These estimated temperatures, along with the column density of the ions, can be used to estimate the abundance ratios of the elements. This abundance ratio can then be used to compare with the solar abundance ratio of the respective elements to check the solar/non-solar mixture of the abundance ratios of the elements. Since we have only upper limits on temperature using Ne and N ions, we assume that $N_{\rm Ne} \geq N_{\rm \neix}$ and $N_{\rm N} \geq N_{\rm \nvi}$. Therefore, we get a lower limit on [N/O] and [Ne/O] abundance ratios. The abundance ratio for [N/O] is $\geq -0.29^{+0.17}_{-0.19}$, [Ne/O] is $\geq 0.37^{+0.14}_{-0.20}$. These values clearly show that Neon is super-solar relative to Oxygen, while N is in solar mixture with O, assuming they exist in the same phase. These estimates clearly show the complex metal mixing and different temperature phases of the CGM of the Milky\,Way.

The $b$ parameter obtained for \ovii is $48.8^{+4.6}_{-12.5}$ km s$^{-1}$ that correspond to a temperature of $2.3^{+0.4}_{-1.0} \times 10^6$ K assuming no non-thermal broadening. This temperature estimate is consistent with the temperature obtained from the column density ratios of \oviii and \ovii within $1\sigma$ error.

\subsection{\texttt{PHASE} model}
\label{sec:phase}

To model all the detected and non-detected absorption lines self-consistently using the Voigt profiles, we fit the observed spectra with the \texttt{PHASE} model (\citealt{Krongold2003,Nicastro2018,Das2019a,Das2021}). \texttt{PHASE} model is a hybrid-ionization model based on collisional ionization and photo-ionization. The parameters of the \texttt{PHASE} model are temperature, equivalent hydrogen column density, the relative abundance of elements, non-thermal broadening, photo-ionization parameter ($U$), and redshift (equivalent to the line of sight velocity that shifts the line center). The abundance of elements is set to a solar value by default, but can be varied independently for He, C, N, O, Ne, Mg, Al, Si, S, Ar, Ca, Fe, and Ni. Therefore, the \texttt{PHASE} model can be used to disentangle various temperature phases and chemical composition of the CGM based on the observed absorption lines. 

To fit with the \texttt{PHASE} model, we adopt the same wavelength range and continuum for each ion as used to detect it in a model-independent way in Sec. \ref{sec:model_ind_res} and Table \ref{tab:detection} for the respective instruments. We freeze the photo-ionization parameter $U$ to its lowest value of $10^{-3.9}$, assuming photo-ionization is negligible. Since the detected absorption line centers are consistent with the resolution element of the respective instruments, we fix the redshift parameter $z$ to $0$ initially. We assume purely thermal broadening of the absorption lines, as shown by \citealt{Das2024}, that the non-thermal broadening along this sightline is insignificant. 

We use the power-law continuum with the same index and normalization over the specified wavelength range for each ion, as discussed previously, and freeze these parameters for further analysis. The continuum model results in $\chi^2$/dof of $5354.35/5620$. We then add a single temperature \texttt{PHASE} model with solar composition, which improves the fit ($\chi^2$/dof of $5202.63/5618$) but cannot reproduce all the detected ions. 

To check the importance of the non-solar composition of various elements relative to O, we allow the composition of C, N, Ne, Mg, and Fe to vary in a single temperature \texttt{PHASE} model, which results in a significant improvement in the fit with $\chi^2$/dof of $5179.24/5612$ with null-hypothesis probability of $0.03 \%$. It suggests the importance of the non-solar chemical composition. We refer to this virial warm-hot phase as \texttt{PHASE\_A} or $T_2$. 

However, this single temperature phase model with non-solar composition could not reproduce the column density of high ionization species like \mgxii and \sixiv, and under-predicts the \oviii column density. It indicates the presence of an additional, higher-temperature phase. Therefore, we add another temperature \texttt{PHASE} model to fit the data. We do not force any parameters of the two-temperature \texttt{PHASE} model to be the same, except that we tie the relative abundance of [C/O] and [N/O] to be the same in both phases, as the observed ions of C and N are not expected to contribute significantly in the hot phase. 

For the higher temperature phase, we vary the relative abundance of Ne, Si, Mg, and Fe. The two-temperature non-solar \texttt{PHASE} model results in $\chi^2$/dof of $5166.68/5606$ with null-hypothesis probability of $3.4 \%$, which is an improvement over the single-temperature non-solar phase model and successfully produces the \mgxii, \sixiv, and total \oviii column density. We refer to this `super-virial' hot phase as \texttt{PHASE\_B} or $T_3$.

The primary motivation of this work is to detect the hot super-virial phase along with the virial phase, which is successfully recovered using our \texttt{PHASE\_A} and \texttt{PHASE\_B} models discussed above. But these two-temperature phases under-predict the column density of \cv K$\alpha$ and \nvi K$\alpha$ lines. To account for the observed excess column density of \cv and \nvi lines, we add another \texttt{PHASE} model. We refer to this phase as \texttt{PHASE\_C} or $T_1$ or sub-virial phase. We allow the temperature and equivalent H column density to vary, and we tie the C and N abundances in this phase to the virial phase. We fix the other metal abundances to the default solar value in this phase, which resulted in $\chi^2$/dof of $5162.39/5604$ with null-hypothesis probability of $9.8 \%$.
However, the ions tracing the sub-virial phase (\cv and \nvi) are blue-shifted; therefore, we vary the redshift parameter (equivalent to line-of-sight velocity) in the \texttt{PHASE\_C}. This improved the fit-statistics further with $\chi^2$/dof of $5156.53/5603$ with null-hypothesis probability of $1.2 \%$.
This additional phase successfully reproduces the excess \cv K$\alpha$ and \nvi K$\alpha$ column density. Therefore, all the observed column density values are well reproduced by our final best-fit \texttt{PHASE} model: \texttt{PHASE\_A}*\texttt{PHASE\_B}*\texttt{PHASE\_C}. We show the results of our best-fit \texttt{PHASE} model in Table \ref{tab:phase} with $1\sigma$ uncertainties.

Since we did not detect \mgxi, which traces the warm-hot gas, and \nex, which traces the hot gas, do we require a non-solar composition of Mg in the warm-hot phase and Ne in the hot phase? To check the significance of non-solar composition of [Mg/O] in warm-hot phase, we freeze all the parameters in our best-fit \texttt{PHASE} model except T, $N_H$, and Mg composition in warm-hot phase, with $\chi^2$/dof of $5156.53/5617$. We then freeze the [Mg/O] composition to solar value, which resulted in $\chi^2$/dof of $5156.55/5618$. We calculated F-test statistics and found the null hypothesis probability of $88 \%$. Similarly, we carried out the above analysis for Ne in the hot phase, which resulted in a null hypothesis probability of $38\%$. These results indicate that the non-solar compositions of Mg in the warm-hot phase and Ne in the hot phase are statistically not required.

Further, we carry out statistical tests to verify if the line of sight velocity is consistent with $0$ km s$^{-1}$ in both the hot and warm-hot phases. To do so, we freeze all the parameters except $T$ and $N_{\rm H}$ in the respective test phase. We then vary the redshift parameter (equivalent to line of sight velocity) in the respective phase and calculate the significance of non-zero line of sight velocity. For the hot phase, we obtain $\chi^2$/dof of $5155.77/5617$ with null-hypothesis probability of $36 \%$. Similarly, for the warm-hot phase, we obtain $\chi^2$/dof of $5154.82/5617$, with null-hypothesis probability of $17 \%$. Therefore, the line-of-sight velocity is consistent with zero in both warm-hot and hot phases.

\begin{table}
    \centering
    \setlength{\tabcolsep}{1pt}
   \renewcommand{\arraystretch}{1}
    \begin{tabular}{|c|c|c|c|}
    \hline
      Parameter & \texttt{PHASE\_C} ($T_1$) & \texttt{PHASE\_A} ($T_2$) & \texttt{PHASE\_B} ($T_3$) \\
      \hline
      T [K] & $2.2\pm 0.5\times 10^5$ & $1.8^{+0.3}_{-0.2} \times 10^6$ & $5.4^{+1.9}_{-0.8} \times 10^7$  \\
    
      $N_{\rm H}$ [cm$^{-2}$] & $1.9^{+0.9}_{-0.8}\times 10^{18}$ & $8.8^{+2.2}_{-4.8} \times 10^{18}$ & $2.5^{+0.7}_{-1.0} \times 10^{21}$  \\ 

      v$_{\rm los}$ [km s$^{-1}$] & \hbox{--} $95^{+23}_{-9}$ & \hbox{--} & \hbox{--} \\
      
      [C/O] & tied to \texttt{PHASE\_A}  & $0.23^{+0.24}_{-0.16}$ & tied to \texttt{PHASE\_A}  \\ 
      
      [N/O] & tied to \texttt{PHASE\_A} & $0.35^{+0.30}_{-0.42}$ & tied to \texttt{PHASE\_A}  \\
      
      [Ne/O] & \hbox{--} & $1.02^{+0.28}_{-0.27}$ & \hbox{--} \\
    
      [Mg/O] & \hbox{--} & \hbox{--} & $0.43^{+0.21}_{-0.29}$  \\
    
      [Si/O] & \hbox{--} & \hbox{--} & $0.93^{+0.22}_{-0.40}$  \\
     
      [O/Fe] & \hbox{--} & $>- 0.30$ & $>1.0$  \\
     
      [Ne/Fe] & \hbox{--} & $>0.72$ & \hbox{--} \\
    
      [Mg/Fe] & \hbox{--} & \hbox{--} & $>1.43$ \\
   
      [Si/Fe] & \hbox{--} & \hbox{--} & $>1.93$ \\
      \hline
    \end{tabular}
    \caption{Best fit \texttt{PHASE} model parameters.}
    \label{tab:phase}
\end{table}

\section{Results}
\label{sec:results}

Using the \texttt{PHASE} model analysis, we infer three temperature phases of the CGM using the X-ray absorption lines towards PKS\,2155-304. As expected from the model-independent temperature results, we detect virial, super-virial, and sub-virial phases from the \texttt{PHASE} model. In Table \ref{tab:phase}, we show the results of our best-fit \texttt{PHASE} model with $1\sigma$ error. 
In Table \ref{tab:detection}, the first column shows the ion, the second column shows the transition, the third column shows the instrument used to detect it, the fourth column shows the equivalent width in m{\r A}, the ninth column shows the column density computed assuming linear part of the curve of growth, the fifth, sixth, seventh, and eighth columns show the column density predicted from \texttt{PHASE\_C} ($N_{T_1}$), \texttt{PHASE\_A} ($N_{T_2}$), \texttt{PHASE\_B} ($N_{T_3}$) and total from all the three phases respectively. 
Therefore, the eighth and ninth columns can be compared with each other. The total column density predicted from the \texttt{PHASE} model is consistent with detection within $1\sigma$ error bars and $3\sigma$ upper limits for non-detections. For \ovii, we detected both the K$\alpha$ and K$\beta$ lines. The ninth column for \ovii in Table \ref{tab:detection} shows the saturation corrected column density for \ovii (line through the column density value). The saturation corrected column density cannot be used for analysis as both the warm-hot and warm phases contribute to the column density of \ovii.

\begin{figure*}
    \centering
    \includegraphics[width=0.33\textwidth]{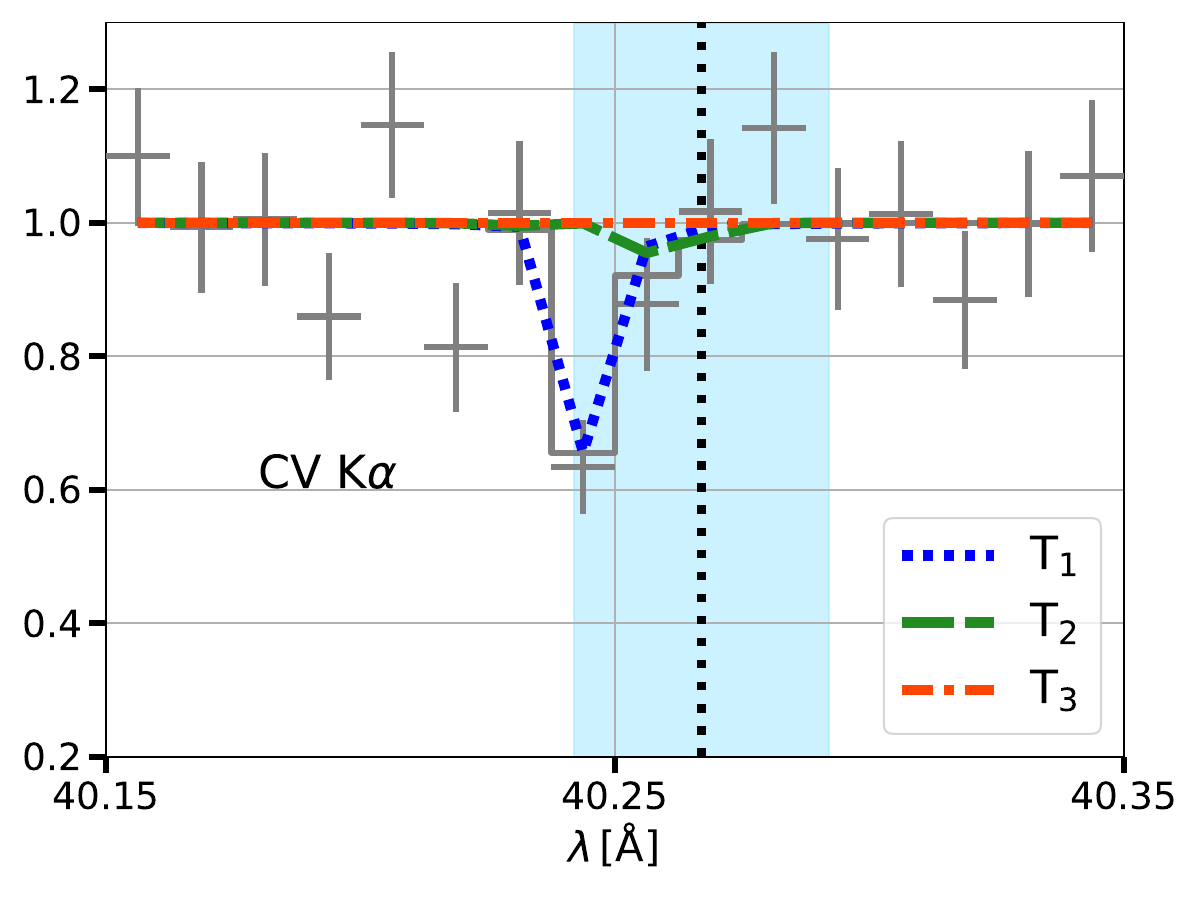}
    \includegraphics[width=0.33\textwidth]{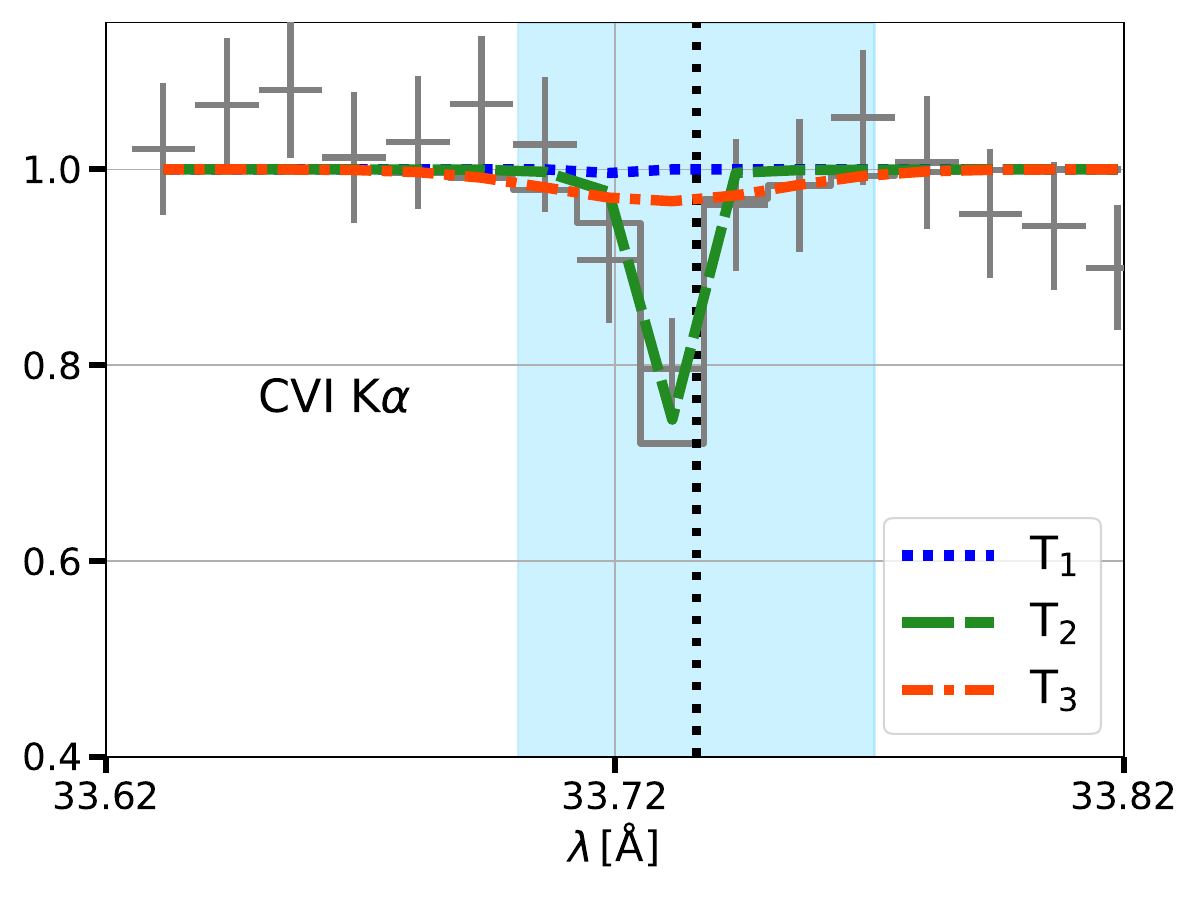}
    \includegraphics[width=0.33\textwidth]{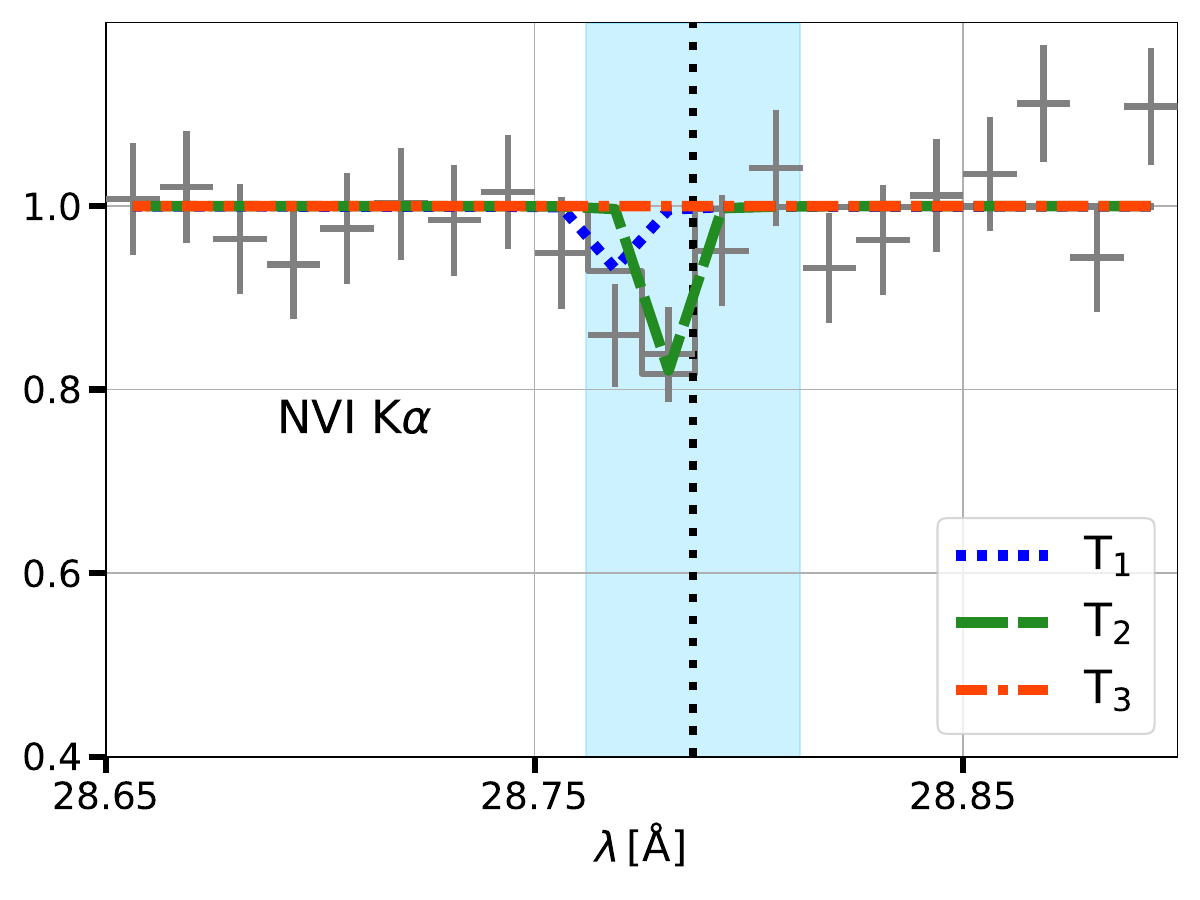}
    \includegraphics[width=0.33\textwidth]{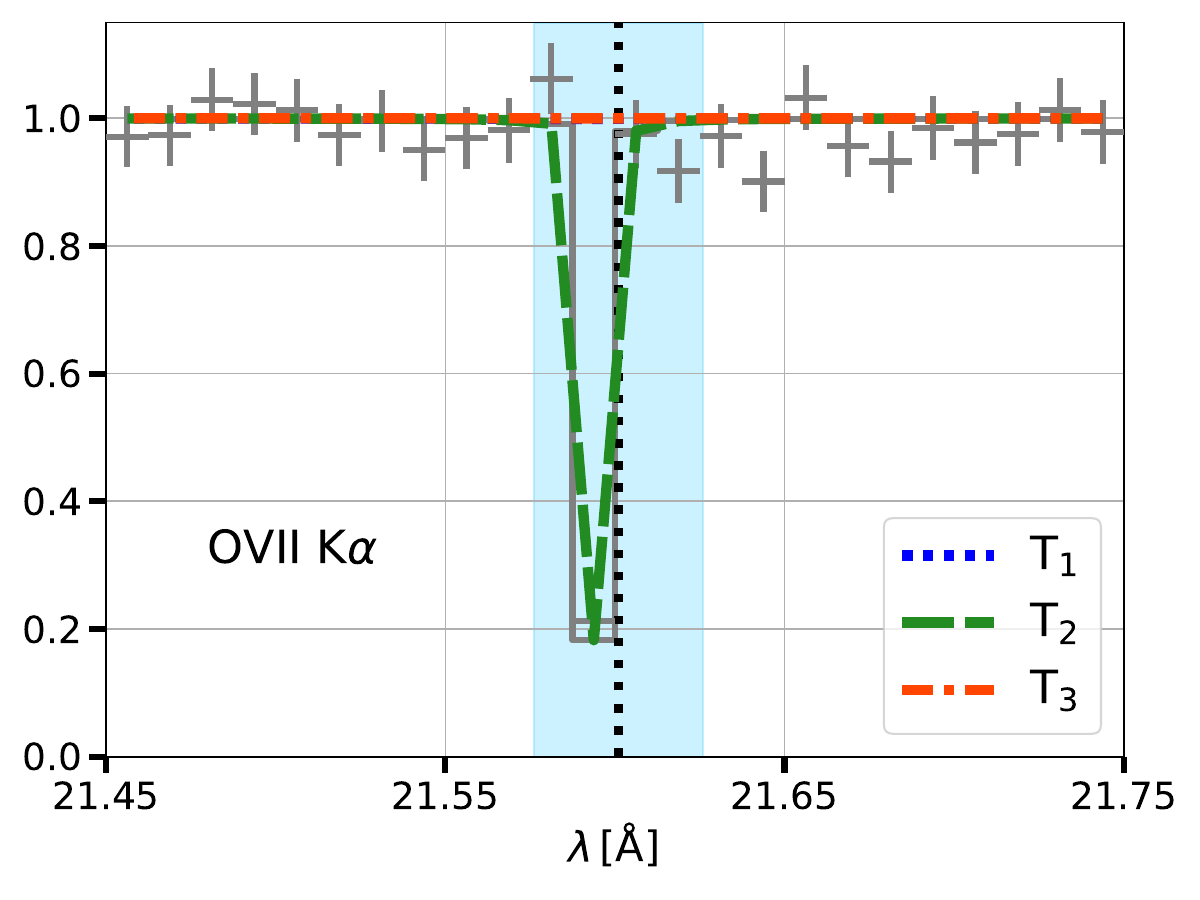}
    \includegraphics[width=0.33\textwidth]{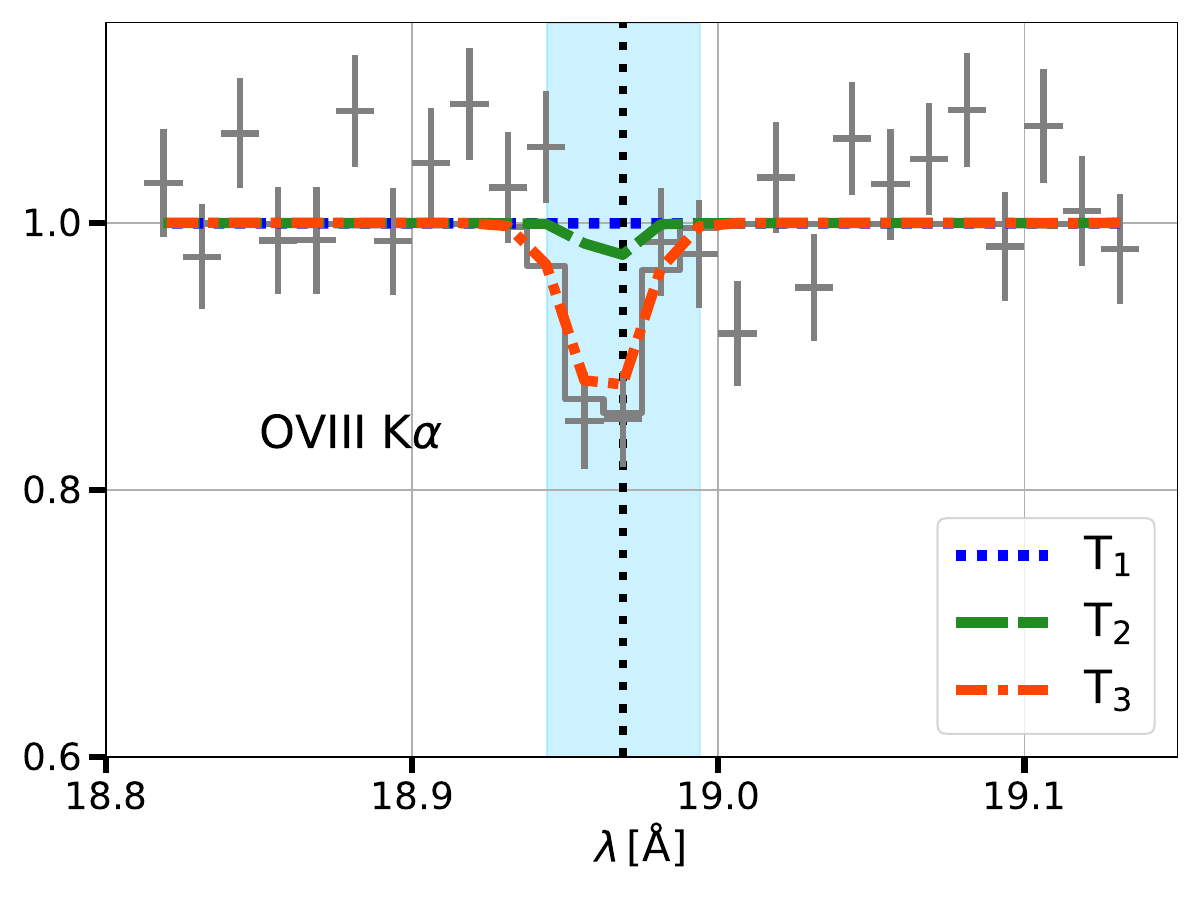}
    \includegraphics[width=0.33\textwidth]{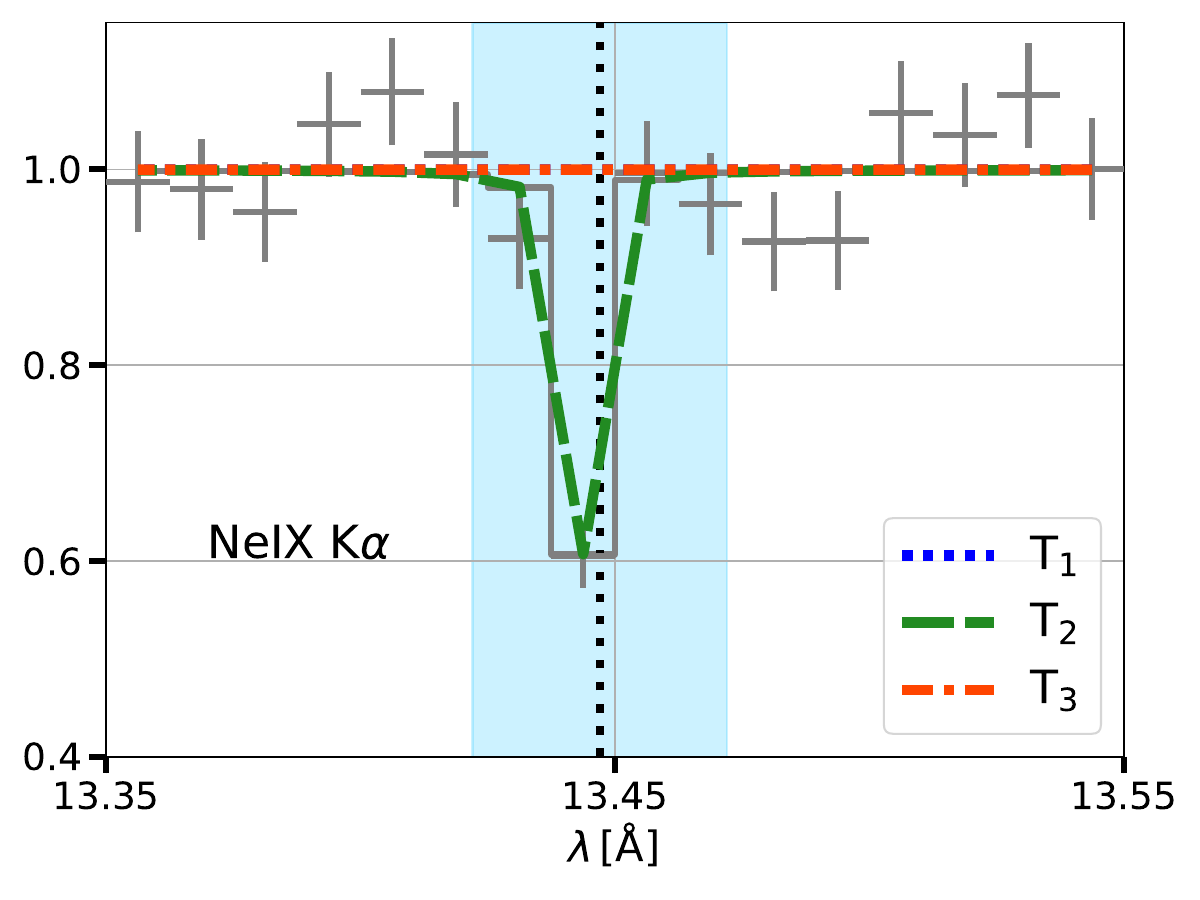}
    \includegraphics[width=0.33\textwidth]{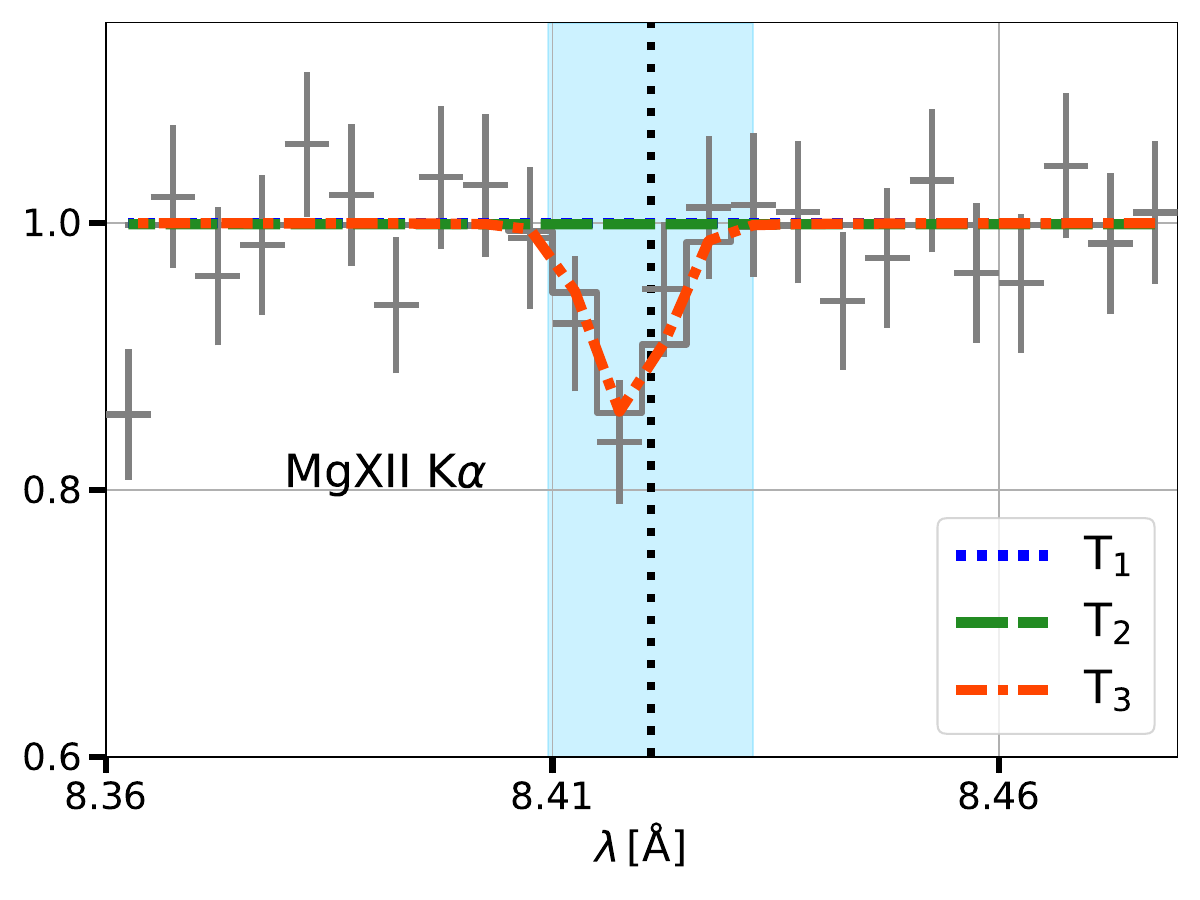}
    \includegraphics[width=0.33\textwidth]{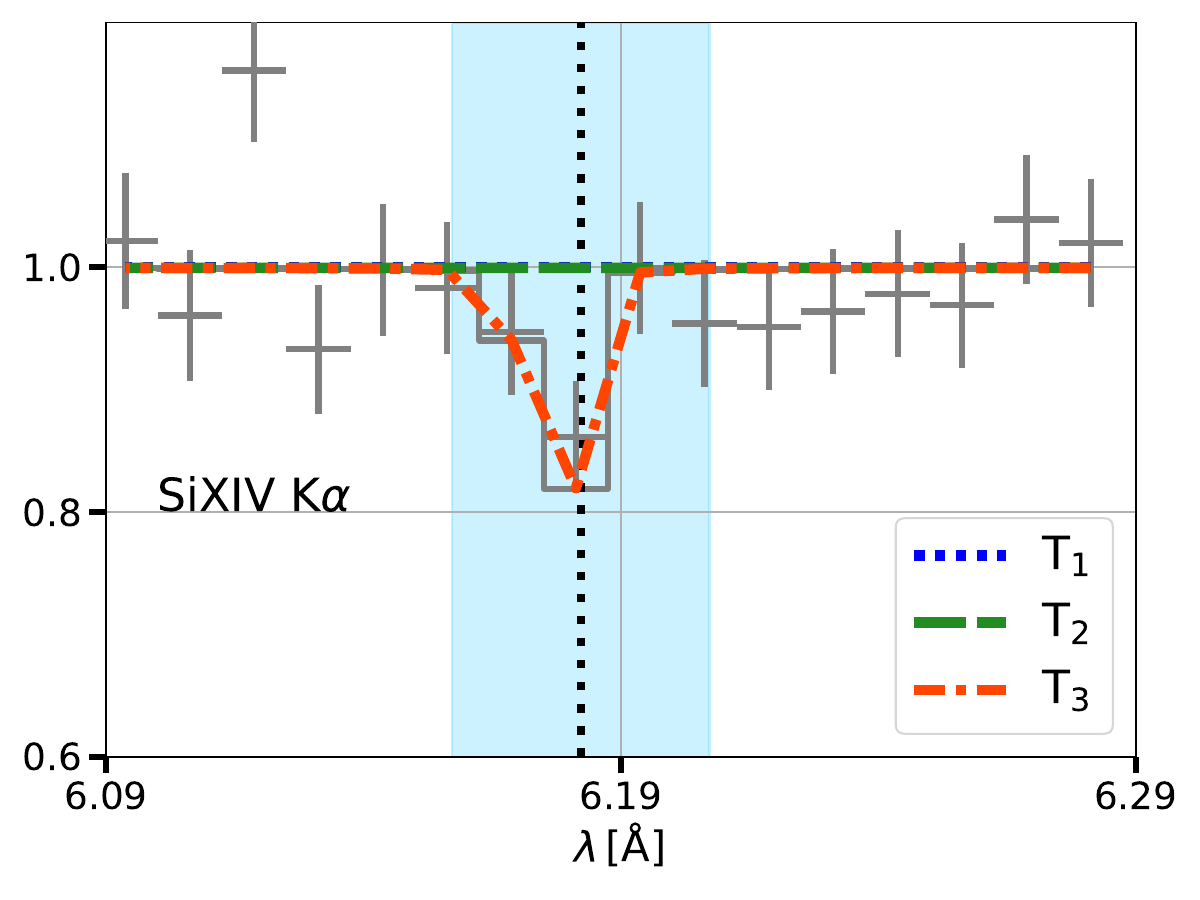}
    \caption{Individual contribution of different temperature phases for detected ions. The grey color shows the best-fit \texttt{PHASE} model, while blue, green, and red colors show the contribution of sub-virial ($T_1$), virial ($T_2$), and super-virial ($T_3$) phases, respectively.}
    \label{fig:phase_contri}
\end{figure*}

In Figure\,\ref{fig:phase_contri}, we show the contribution of different phases for all the detected ions from our best-fit \texttt{PHASE} model. The blue dotted, green dashed, and red dashed-dotted lines show the contribution of warm ($T_1$), warm-hot ($T_2$), and hot ($T_3$) phases, respectively. 
Our analysis indicates that \mgxii and \sixiv originate exclusively from the hot phase, whereas \neix comes from the warm-hot phase. \oviii has a significant contribution from both warm-hot and hot gas, while \ovii has contributions from warm-hot and warm phases. This suggests that the observed \ovii and \oviii column density ratios cannot be used exclusively to estimate the temperature of the gas in the CGM, as they contribute significantly to different temperature phases. \cvi comes from the warm-hot phase with a contribution from the hot phase, and \nvi comes from both the warm and warm-hot phase. \cv has a predominant contribution from the warm phase (see Table \ref{tab:detection}). Therefore, a similar conclusion holds for estimating temperature using the column density ratios of \cvi and \cv.

\subsection{Temperature}

For the virial, super-virial, and sub-virial phases, we obtain the best-fit temperature of $1.8^{+0.3}_{-0.2} \times 10^6$ K, $5.4^{+1.9}_{-0.8} \times 10^7$,K, and $2.2\pm 0.5\times 10^5$\,K, respectively. The best-fit virial temperature along 1ES\,1553+113 \citep{Das2019a}, Mrk\,421 \citep{Das2021}, and NGC\,3783 \citep{McClain2023} is  $1.29^{+0.16}_{-0.27}\times 10^6$\,K, $1.5 \pm 0.06 \times 10^6$\,K, $6.76^{+2.79}_{-1.01}\times 10^5$\,K, while the super-virial temperature is $1.15\pm 0.26 \times 10^7$\,K, $3.2^{+0.94}_{-0.31} \times 10^7$\,K, and $4.07^{+1.3}_{-0.52}\times 10^6$\,K respectively. The sub-virial temperature along Mrk\,421 is $3.0\pm 0.25 \times 10^5$\,K. The best-fit virial, super-virial, and sub-virial temperatures for the stacked sightlines \citep{Lara-DI2024a} are $1.55^{+0.23}_{-0.26} \times 10^6$, $3.16^{+0.23}_{-0.21} \times 10^7$, and $2.45^{+0.86}_{-0.36} \times 10^5$\,K, respectively.

\begin{figure*}
\centering
\includegraphics[width=\textwidth]{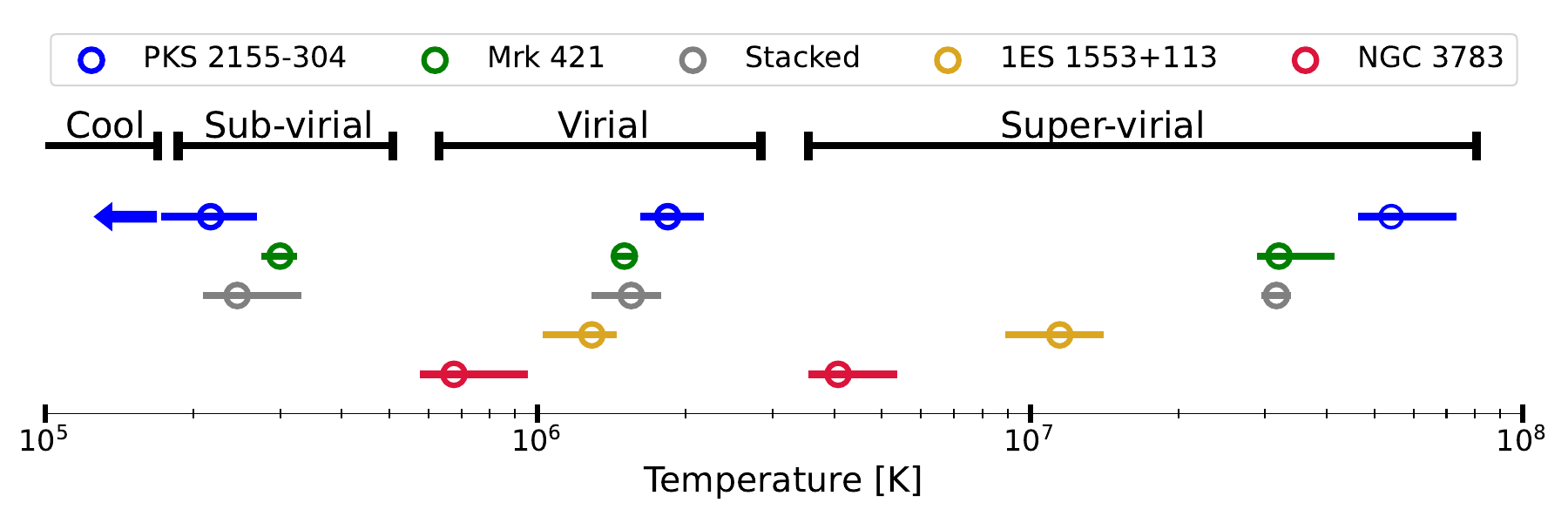}
\caption{Comparison of best-fit temperatures and associated $1\sigma$ uncertainties from previous studies along various lines of sight (Mrk\,421, green, \citealt{Das2021}; stacked, grey, \citealt{Lara-DI2024a}; 1ES\,1553+113, yellow, \citealt{Das2019a};  NGC\,3783, red, \citealt{McClain2023}). The temperature ranges from our study (PKS\,2155-304) are plotted in blue. The left arrow shows the cool gas temperature estimated from the column density ratios of \ov (non-detections) and \oiv (detections) in our study. Therefore, we detect four temperature phases along PKS\,2155-304. This is the first time that $4$ temperature phases are inferred in such studies using X-ray absorption lines. This plot shows the multiphase nature of CGM along various lines of sight and within specific phases as well.}
\label{fig:temp_compare}
\end{figure*}

Figure\,\ref{fig:temp_compare} shows the best-fit temperature and $1\sigma$ uncertainties from previous studies along various lines of sight (Mrk\,421, green; stacked, grey; 1ES\,1553+113, yellow; NGC\,3783, red). The temperature ranges from this work (along PKS\,2155-304) are plotted in blue. The left arrow shows the cool phase temperature estimated using \ov (non-detection) and \oiv (detection) line ratios. This marks the first time that cool gas has been detected via X‑ray absorption lines in such studies. The three points in each row (same color) show the temperature for the sub-virial, virial, and super-virial phases along PKS\,2155-304, Mrk\,421, and stacked sightlines, while along 1ES\,1553+113 and NGC\,3783, the two points show virial and super-virial phases. 
For a given sightline, the temperatures of different phases show no overlap, considering the $1\sigma$ variation, which points to the robust detection of these distinct phases and a temperature valley in between. There is no overlap between the hot and warm-hot temperatures from all the observations within $1\sigma$ uncertainties. This shows the clear demarcation between virial and super-virial phases. 
Similarly, the sub-virial and virial phases are distinct, as there is no overlap between the temperatures of these two phases within $1\sigma$ error. The temperature values in the sub-virial phase are consistent with each other within $1\sigma$ uncertainty. This temperature comparison plot indicates the multiphase nature of CGM along all directions.

We detected four distinct temperature phases along PKS\,2155-304 for the first time. Along Mrk\,421, \citealt{Das2021} found three temperature phases. Since \citealt{Das2021} did not look for \oiv and \ov X-ray absorption lines, a cooler temperature gas may also be present along Mrk\,421. Similarly, a cooler temperature gas may also exist in stacked sightlines \citep{Lara-DI2024a}. 
Along 1ES\,1553+113, \citealt{Das2019a} detected warm-hot and hot gas using RGS1/RGS2 in \emph{XMM-Newton}. However, they could not detect a lower temperature gas because of the limited wavelength coverage of RGS1/RGS2. Their hot gas temperature is also lower than the hot gas temperature among other sightlines (except NGC\,3783). This is again because of the limited wavelength coverage of RGS1/RGS2 at lower wavelengths. They detected \nex as a tracer of hot gas. Other hot gas tracer ions like \mgxii and \sixiv, which fall at lower wavelengths, might be present along this sightline, which can further modify the hot gas temperature to higher values, but could not be detected. 

The warm-hot gas temperature along NGC\,3783 is the lowest among other sightlines. This could be the weighted average temperature of warm-hot and warm gas. \citealt{McClain2023} only used MEG for their study and did not look for ions tracing the lower temperature gas due to the limited wavelength coverage of MEG. As a hot gas tracer, they only detected \nex and could not detect other ions. The limited number of detected ions for warm-hot and hot gas may not reflect the true temperature distribution along a given sightline, which can result in average temperatures between two temperature phases. This could be the case along NGC\,3783. 
A large range in the inferred temperature along various sightlines also points towards the temperature inhomogeneities in the CGM. The inhomogeneity is larger for the hot phase, which could be the result of scattered stellar outflows from the Galactic disk.

\subsection{Equivalent H column density}

Virial, super-virial, and sub-virial phases have an equivalent H column density (relative to solar abundance of Oxygen in that phase) of $8.8^{+2.2}_{-4.8} \times 10^{18}$, $2.5^{+0.7}_{-1.0} \times 10^{21}$, and $1.9^{+0.9}_{-0.8} \times 10^{18}$ cm$^{-2}$ respectively. Equivalent H column density towards 1ES\,1553+113, Mrk\,421, and NGC\,3783 for virial phase is $13.3^{+5.0}_{-4.5} \times 10^{18}$, $8.4\pm 0.8 \times 10^{18}$, $1.26^{+0.83}_{-0.64} \times 10^{20}$, while for super-virial phase is $2.9^{+2.4}_{-2.0} \times 10^{20}$, $9.1^{+2.7}_{-2.9} \times 10^{20}$, and $4.47^{+2.94}_{-2.28} \times 10^{19}$ cm$^{-2}$. The equivalent H column density for the sub-virial phase towards Mrk\,421 is $1.1^{+0.2}_{-0.3} \times 10^{18}$ cm$^{-2}$.
Note that the uncertainties along Mrk\,421 and 1ES\,1553+113 are with $90\%$ and $99.73\%$ confidence levels, respectively, while others are with $1\sigma$.
The temperature and equivalent H column density for the super-virial phase predicted by our best-fit \texttt{PHASE} model are maximum along this line of sight compared to the other three lines of sight (\citealt{Das2019a,Das2021,McClain2023}) and stacked sightlines \citep{Lara-DI2024a}. A large variation in inferred temperature and equivalent H column density for the hot phase across different sightlines suggests the inhomogeneous nature of this phase. 

\subsection{Abundance Ratios}

\begin{figure}
    \centering
    \includegraphics[width=0.8\columnwidth]{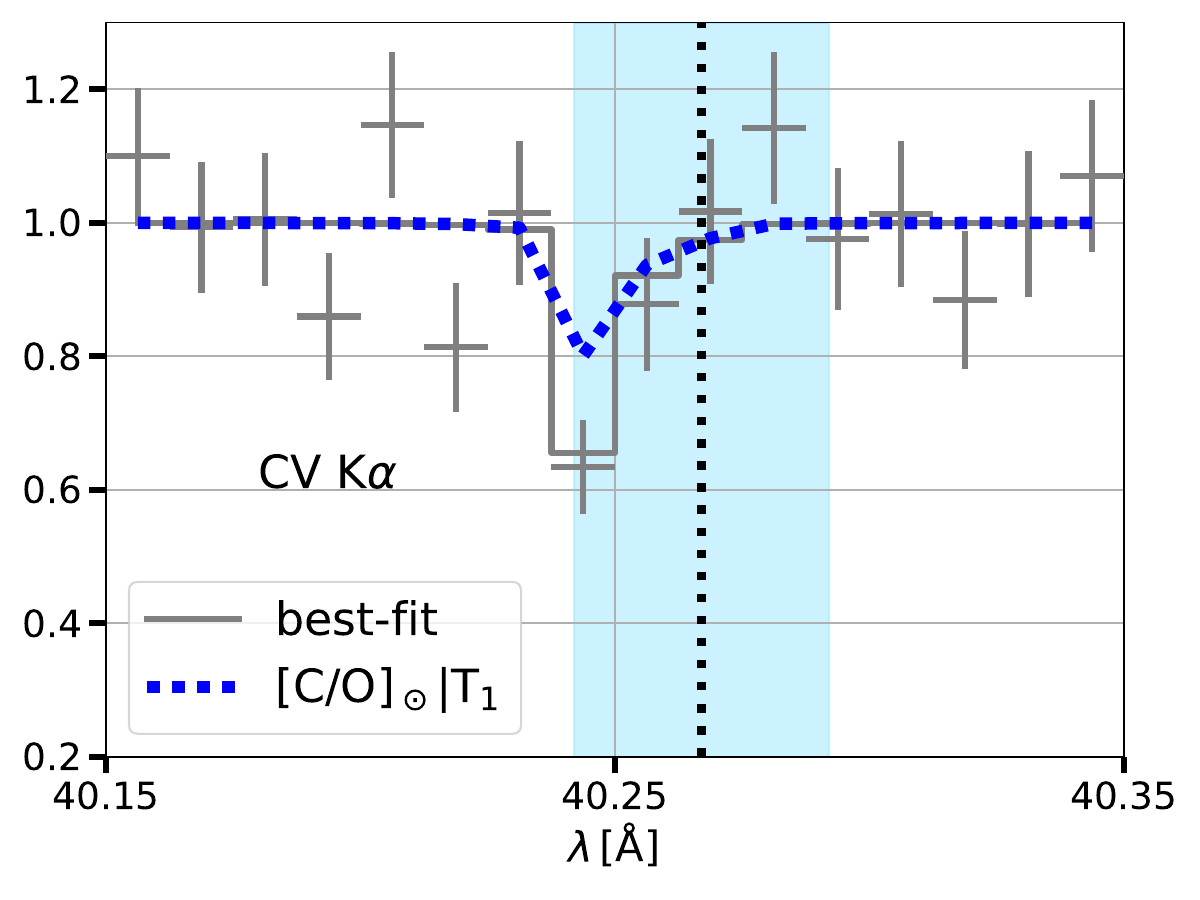}
    \includegraphics[width=0.8\columnwidth]{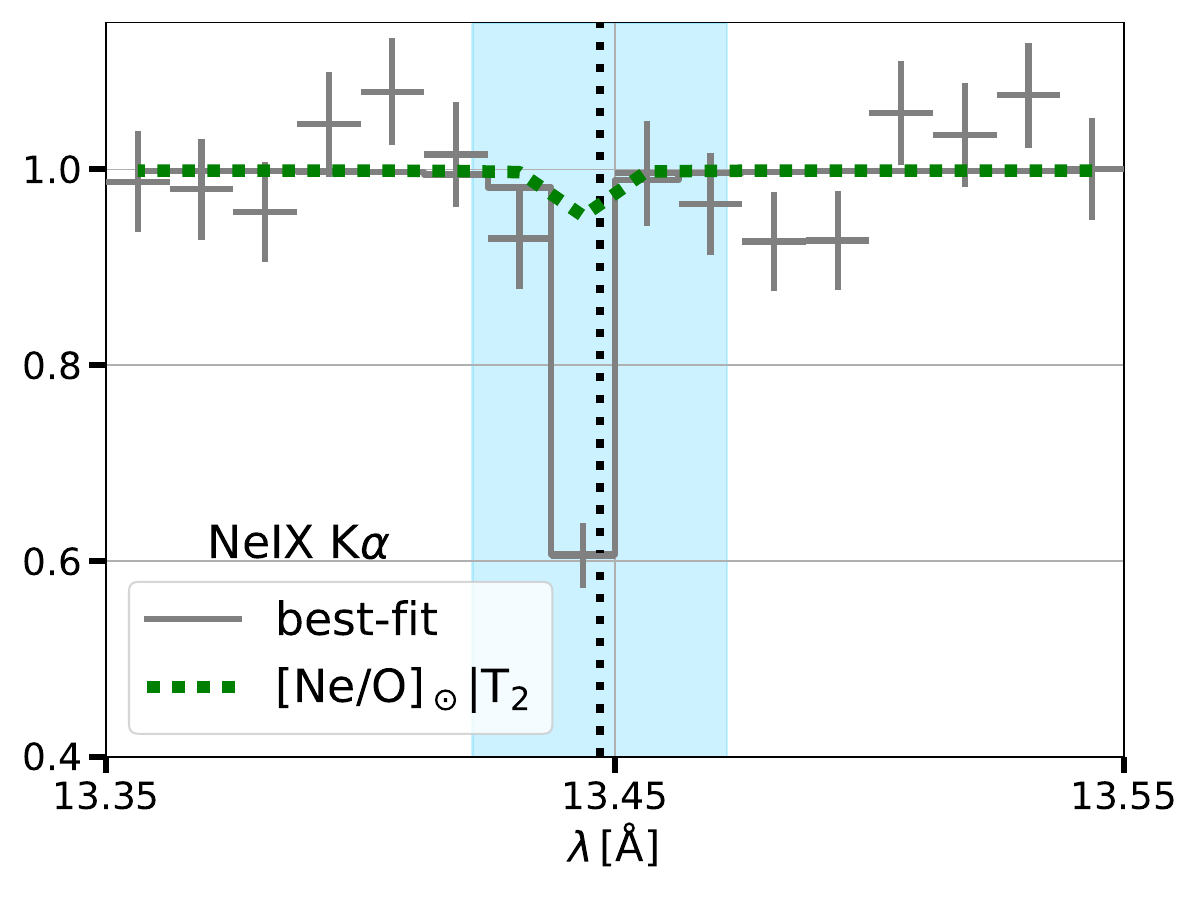}
     \includegraphics[width=0.8\columnwidth]{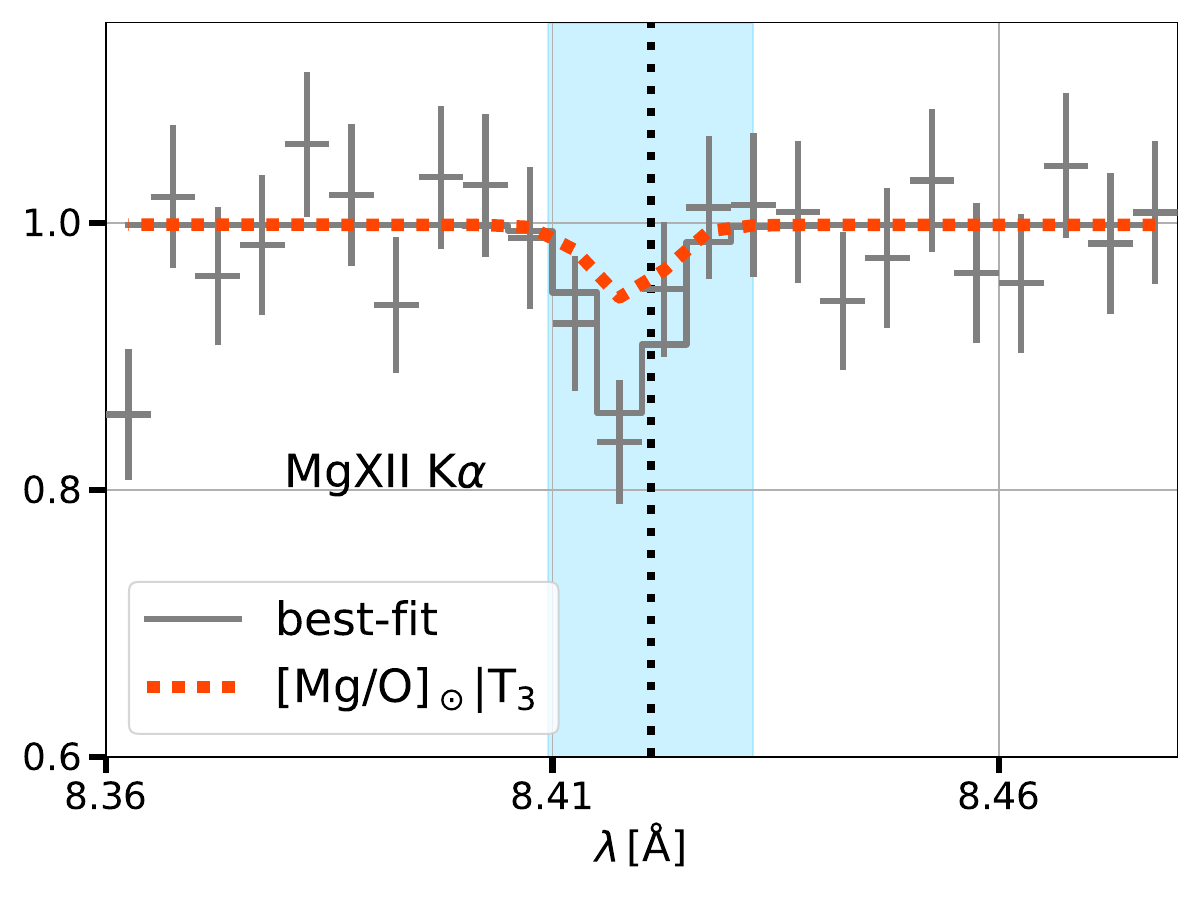}
      \includegraphics[width=0.8\columnwidth]{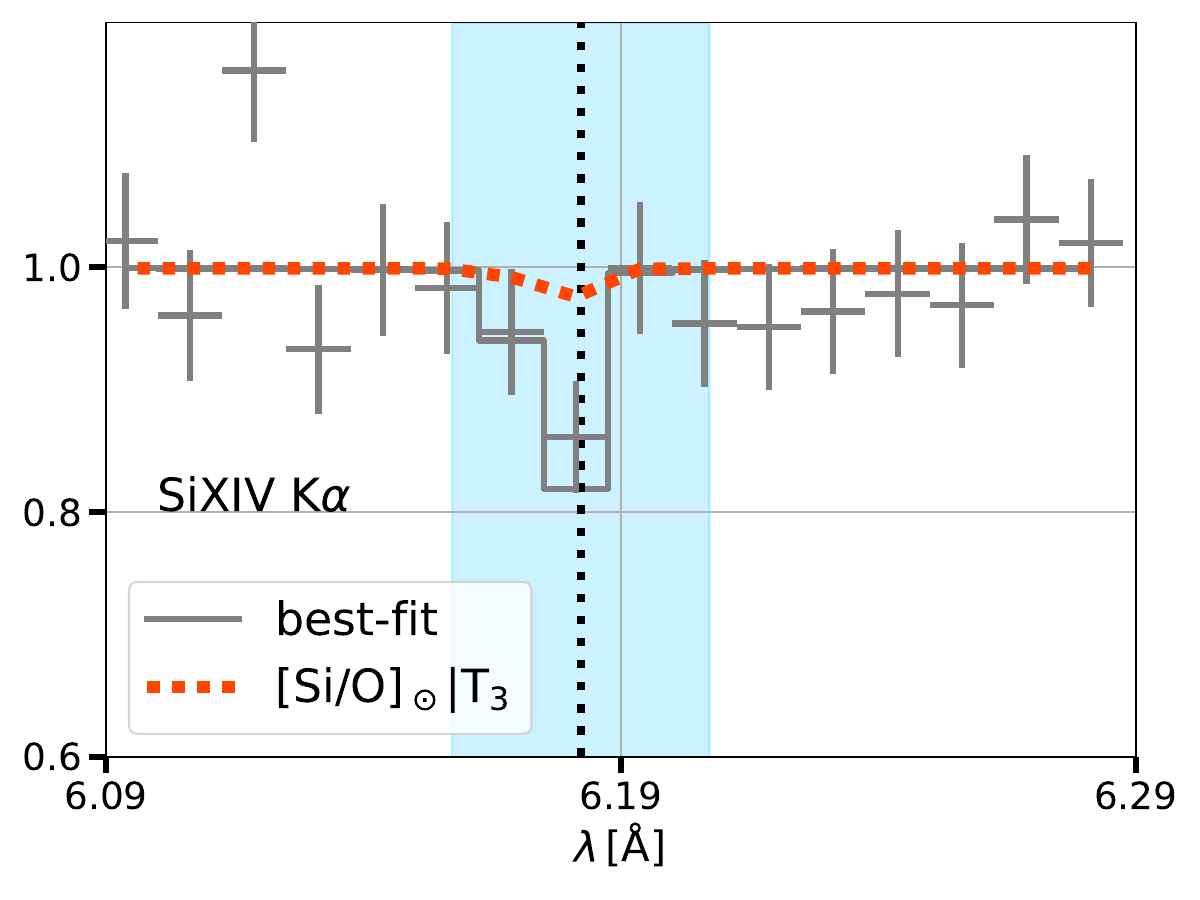}
      \caption{Non-solar abundance ratios of C, Ne, Mg, and Si relative to O in the respective phases. }
      \label{fig:non-solar-composition}     
\end{figure}

We find that the abundance ratios of the ions in the warm-hot and hot phases are non-solar. C and Ne are super-solar relative to Oxygen in the virial phase. Mg and Si are super-solar relative to Oxygen in the super-virial phase. In Fig. \ref{fig:non-solar-composition}, we show the importance of non-solar composition of the abundance ratio of ions. The dotted colored line in sub-plots shows the normalized absorption profile of various ions for solar abundance ratios of the respective element relative to O in various phases.

\begin{figure}
    \centering
    \includegraphics[width=\columnwidth]{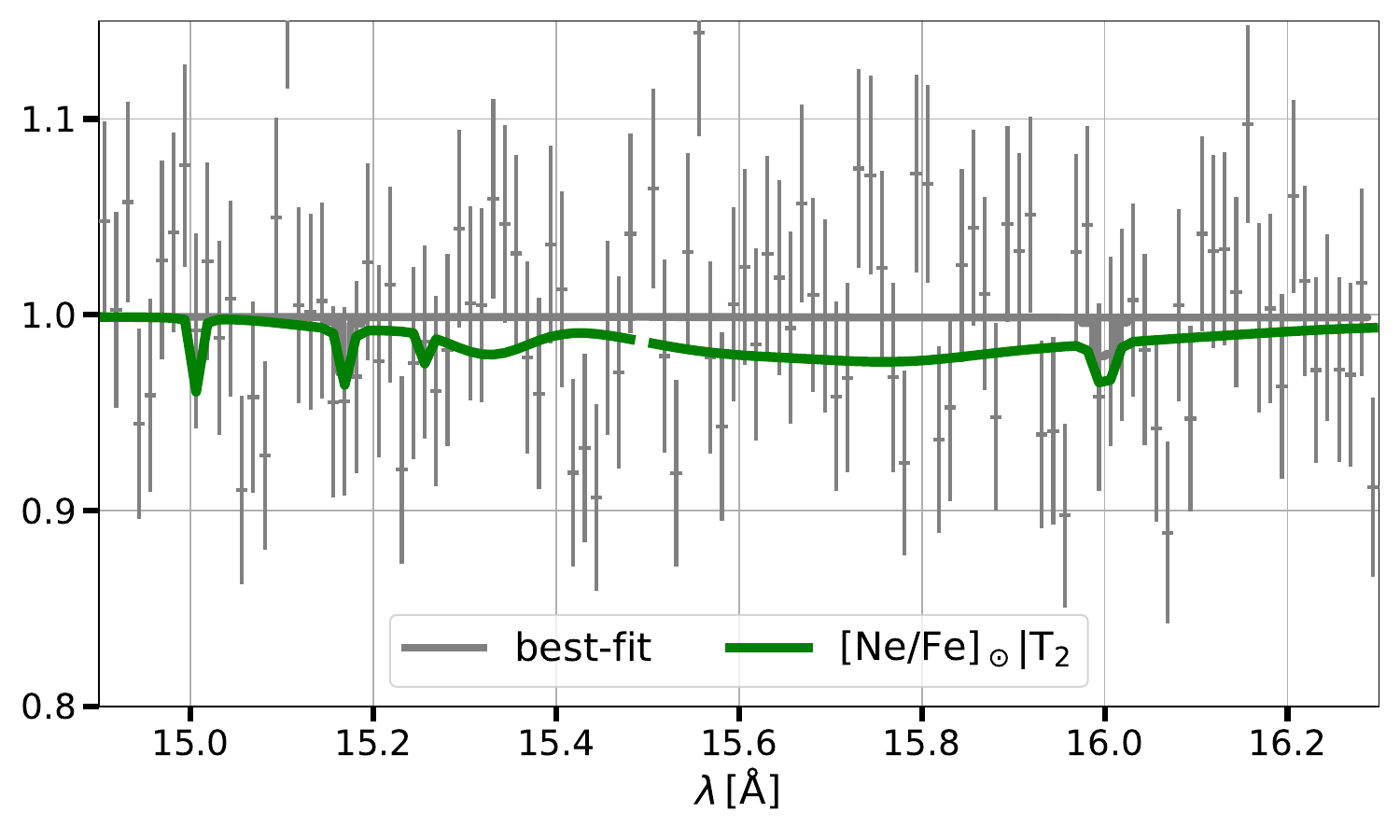}
    \includegraphics[width=\columnwidth]{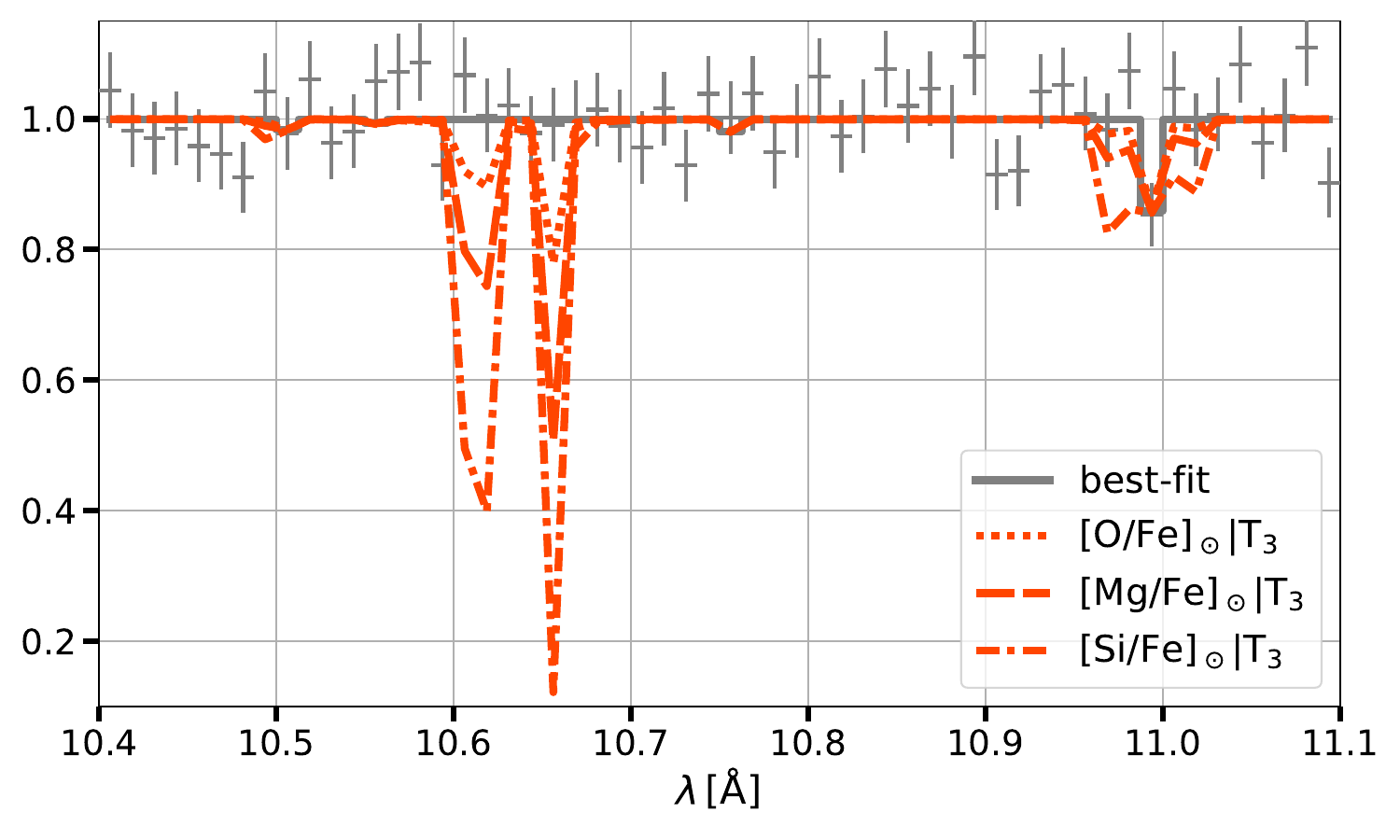}
    \caption{\textit{Top panel}: The green line shows the normalized spectrum for the Solar abundance ratio of Ne to Fe in the warm-hot phase. \textit{Bottom panel}: The dotted, dashed, and dashed-dotted red lines show the normalized spectrum for solar abundance ratios of O to Fe, Mg to Fe, and Si to Fe in the hot phase. This shows the $\alpha-$enhancement in virial and super-virial phases.}
    \label{fig:alpha_abundance}   
\end{figure}

Along with these elements, we have also varied the composition of Iron in both virial and super-virial phases. We find $\alpha$-enhancement in both the virial and super-virial phases, which is in line with all the previous studies along different sightlines. In the top panel of Fig. \ref{fig:alpha_abundance}, the green line shows the normalized spectrum for [Ne/Fe]$_\odot$ in the virial phase. In the bottom panel, the dotted, dashed, and dashed-dotted red lines show the normalized spectrum for [O/Fe]$_\odot$, [Mg/Fe]$_\odot$, and [Si/Fe]$_\odot$, respectively, in the hot phase. The virial and super-virial phases are $\alpha$-enhanced as evident from this plot.

\section{Discussion}
\label{sec:discussion}

\subsection{Comparison with previous studies}

\subsubsection{X-ray}

\citealt{Nevalainen2017} studied and analyzed the $z=0$ absorption lines in the spectra of PKS\,2155-304. They detected various ions individually in the spectra of RGS1, RGS2, LETG/HRC, and LETG/ACIS (see their Figure\,2). Since their motivation was to study the X-ray counterpart of the far-ultraviolet observations, which revealed a transition temperature along this sightline, they did not look for the ions tracing very hot gas in the CGM. Therefore, they did not analyze the MEG/ACIS data, which is sensitive to detecting such ionic transitions. They have reported a detection of the \ov line in RGS1 and LETG/HRC; however, we did not detect it (see Figure\,\ref{fig:detection}) in any of the instruments. The $3\sigma$ and $2\sigma$ upper limits on the EW from our analysis are $5.6$ and $4.3$ m{\r A} respectively, which is consistent with their detection of $3.0\pm 1.5$ m{\r A} for RGS1. However, the number of observations used in our analysis is larger than those analyzed in \citealt{Nevalainen2017}. Our analysis has a total exposure time of $1.98$ Ms, compared to $1.7$ Ms in \citealt{Nevalainen2017} for RGS1.

A similar study is carried out along various other lines of sight, including 1ES\,1553+113 (\citealt{Das2019b}), Mrk\,421 (\citealt{Das2021}), and NGC\,3783 (\citealt{McClain2023}). This is the first study that is conducted in the southern latitude ($b=-52.2^{\circ}$), which indicates the presence of super-virial in the southern Galactic latitude as well.
Mrk\,421 ($l=179.83^{\circ}$) lies in the Galactic anti-center direction, while PKS\,2155-304 is towards the Galactic center ($l=17.7^{\circ}$). \citealt{Das2021} inferred three temperature phases along Mrk\,421, which is similar to three temperature phases (hot/warm-hot/warm) along this sightline. The temperature, equivalent H column density, and abundance pattern are similar along both lines of sight, even though they lie completely in opposite directions. This suggests the widespread nature of super-virial gas detected in X-ray absorption studies. Further observations with high signal-to-noise grating data across a broader array of sightlines are necessary to clarify and characterize the properties of the hot gas observed in X-ray absorption studies.

\subsubsection{UV}

\begin{figure}
    \centering
    \includegraphics[width=\linewidth]{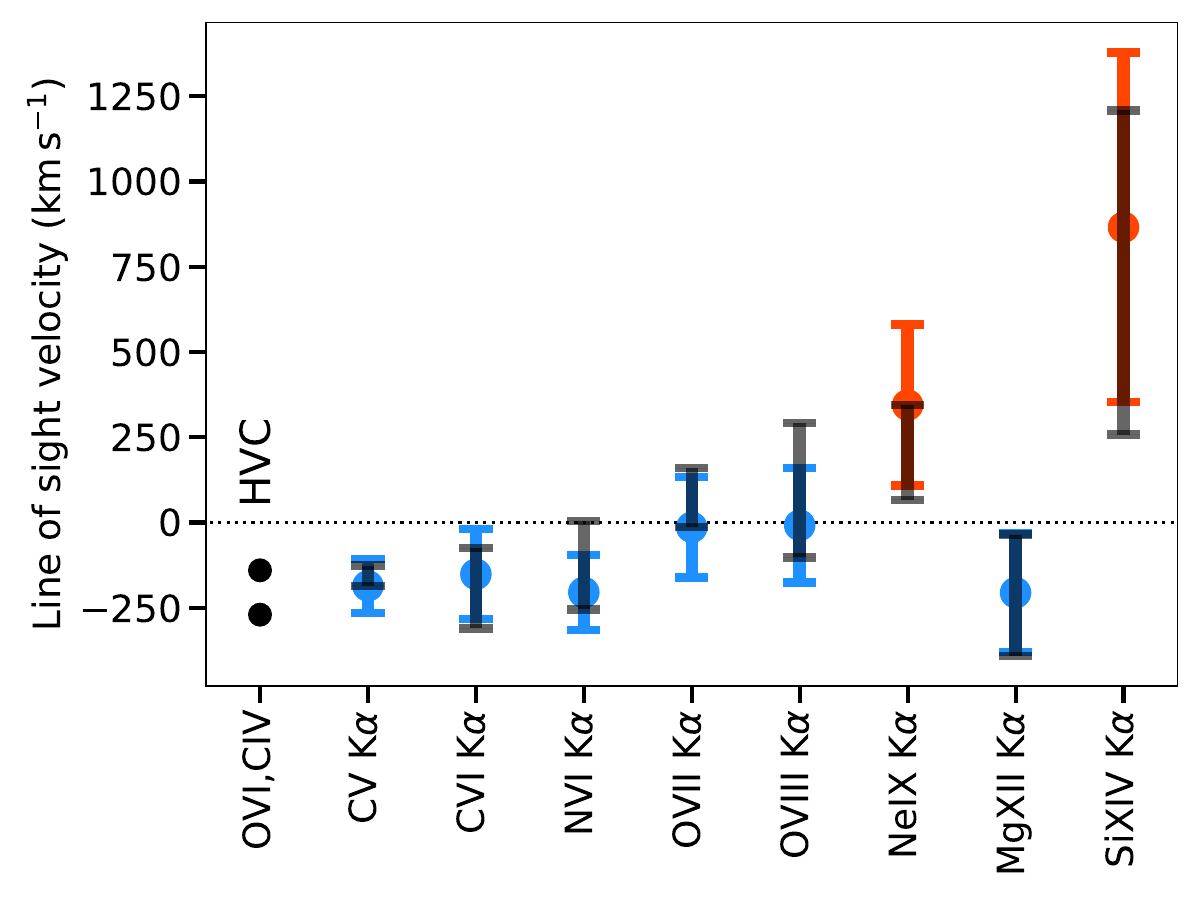}
    \caption{The line of sight velocity\footnote {estimated from the best-fit and $z=0$ wavelength value ($v/c=(\lambda-\lambda_0)/\lambda_0$) as shown in Figure\,\ref{fig:detection}} of the detected transitions. The blue and red colors show blueshift and redshift, respectively. The error bars in the respective color indicate the resolution element of the instruments used to detect these ions relative to the best-fit velocity. The error bars in black show the $1\sigma$ error about the best-fit value. Two black points on the left side of the plot correspond to the two velocity components ($\hbox{-}140$ and $\hbox{-}270$ km s$^{-1}$) detected in UV (\ovi and \civ) along this sightline for HVC \citep{Collins2004}.}
    \label{fig:vel}
\end{figure}

There is observational evidence of highly ionized High Velocity Clouds (HVC) towards PKS\,2155-304. \citealt{Collins2004} used the Hubble Space Telescope (HST) and the Far Ultraviolet Spectroscopic Explorer (FUSE) to study the UV absorption towards PKS\,2155-304. They detected two absorbing components at $V_{LSR}$ of $-140$ and $-270$ km s$^{-1}$ in \ovi and \civ. These negative LSR velocities are also consistent with our blue-shifted detection of ions like \cv, \cvi, and \nvi (see Figure\,\ref{fig:detection} and \ref{fig:vel}), which possibly have a significant contribution from the HVCs.

In Figure\,\ref{fig:vel}, we show the line of sight velocity for each ion calculated from the best fit and theoretically predicted $z=0$ wavelength value. The blue and red colors show blueshift and redshift, respectively, and the error bars in the respective colors show the resolution element around the best-fit values. The error bars in black color show the measured $1\sigma$ error about the best fit value. The two black points on the left side of the Figure\,\ref{fig:vel} correspond to the two velocity components of the UV High Velocity Clouds (HVCs) \citep{Collins2004}. \cv and \nvi lines, which are the tracers of the warm/cool phase, are blue-shifted. \ovii and \oviii lines are roughly consistent with zero velocity as these ions trace the virial gas, which is expected to be in hydrostatic equilibrium with the dark matter Halo. The \sixiv line, which traces hot gas, is red-shifted. It shows that the hot gas is probably outflowing due to feedback from the Galactic disk and Galactic center. The \mgxii line is also expected to be red-shifted as a tracer of the hot gas, but is rather blue-shifted. This shows the complex and inhomogeneous mixing of inflowing and outflowing gas in the CGM. \citealt{McClain2023} also detected the blueshifted and redshifted absorption lines in their analysis towards NGC\,3783. They found \ovii K$\alpha$, \ovii K$\beta$, and \neix K$\alpha$ lines to be redshifted, while \oviii K$\alpha$ and \nex K$\alpha$ lines were blueshifted (see their Figure\,$1$ and Table $1$). However, they did not claim detections of inflow/outflow because the available evidence was scarce and based on a limited set of tracer ions.

Along with various other lines, \citealt{Collins2004} detected the \ovi line ($1031.93$ {\r A}) with column densities of $13.80\pm 0.03$ and $13.56\pm 0.06$ (in log$_{10}$) in the two absorbing components. Since X-ray instruments cannot resolve these two components, we add the two column densities to compare with our analysis. The total \ovi column density is therefore $9.94^{+0.99}_{-0.89} \times 10^{13}$ cm$^{-2}$. Since the \ovi K$\alpha$ ($22.02$ {\r A}) line is highly blended with the \oii K$\beta$ line in X-ray (\citealt{Mathur2017}), we cannot independently compute the \ovi column density from X-ray. However, the \ovi column density predicted from our best-fit \texttt{PHASE} model is $1.06^{+0.57}_{-0.34}\times 10^{14}$ cm$^{-2}$, which is consistent with the UV measurements within $1\sigma$ uncertainties. 

The \ovi column density predicted from X-ray along Mrk\,421 was $1.82^{+0.14}_{-0.27} \times 10^{14}$ cm$^{-2}$ which was smaller by a factor of $\approx 2$ than the observed UV value of $(2.66-3.15) \times 10^{14}$ cm$^{-2}$ along that direction. Even photo-ionization in the warm phase could not account for the excess \ovi observed along this sightline in UV, which suggests the effect of other processes like radiative cooling or non-thermal sources \citep{Das2021} deviating from CIE. However, along PKS\,2155-304, the consistency in \ovi column density between X-ray and UV suggests that the warm phase is in CIE. Thus, the ionization conditions along the two sightlines are different for the warm gas traced by \ovi.

The total \civ ($1548.20$ {\r A}) column density from the UV observation is $7.22^{+0.62}_{-0.55} \times 10^{13}$ cm$^{-2}$. The \civ column density predicted by the \texttt{PHASE} model is $2.85^{+2.68}_{-1.26}\times 10^{13}$ cm$^{-2}$, which is less than the UV measurements. 
The excess \civ is therefore, $N$(\civ)$_{UV}$\hbox{--}$N$(\civ)$_X$ = $4.37^{+2.06}_{-0.71} \times 10^{13}$ cm$^{-2}$. 
This `missing' \civ in X-rays might come from a further lower temperature gas, as suggested by the temperature estimated from the column density ratios of \ov and \oiv (see Figure\,\ref{fig:illustration}). 
The observed UV and inferred X-ray \civ column density along Mrk\,421 were consistent with each other within $2\sigma$, but the X-ray \civ column density was larger than what was observed in UV. The excess column density in that case would come from low/intermediate velocity clouds (LVC/IVC) in the disk/halo.
However, along PKS 2155-304, the \civ UV column density is larger than what is predicted from our X-ray analysis. Therefore, the two lines of sight are also different in \civ properties. 

\subsection{Origin of the super-virial gas}

The X-ray emitting and absorbing super-virial gas is believed to have distinct origins. \citealt{Bhattacharyya2023} showed that the emitting super-virial gas has a dominant contribution from Fe line emission, whereas the absorbing super-virial gas is found to be $\alpha$-enhanced. 
\citealt{Bisht2024b} also showed that the absorbing and emitting super-virial gas has to be treated separately, based on the typical value of Emission Measure (EM) and column density.
The emitting super-virial gas is ubiquitous across the CGM, and \citealt{Bisht2024b} showed that a puffed up disk-like geometry with a typical scale radius and scale height of $\approx 5$, and $1$, respectively, can match the characteristics of this gas. 
\citealt{Bisht2024b} further showed that the disk-wide stellar outflows from the MW central disk can give rise to such a puffed-up disk with super-virial temperatures using hydrodynamical simulations of stellar outflows. However, this puffed-up disk-shaped gas cannot account for the extremely high column density of ions detected in absorption. 

\citealt{Vijayan2022} in their hydrodynamical simulations of star-forming galaxies also found hot gas ($\sim 10^7$ K; see their Figure\,$7$). But the star formation rate (SFR) they considered, $10 \, M_\odot \rm \, yr^{-1}$ is higher than the SFR of MW \citep{Elia2025}. The signatures of super-virial gas are also found by \citealt{Roy2024} as a result of compressive heating of infalling gas into a galaxy in their simulation. However, their temperature values are significantly lower than those inferred from X-ray absorption studies, and they have not quantified the characteristics of super-virial gas (super-solar abundance ratios, $\alpha$-enrichment) beyond temperature.

Based on the properties of the absorbing super-virial gas (high temperature, super-solar abundances, and $\alpha-$enrichment), \citealt{Bisht2024a} showed that the core collapse supernovae from runaway stars in the extra-planar regions along these specific sightlines can give rise to these characteristics around $\sim 100$ kyr timescale. They showed that the reverse shock in the supernova remnant (SNR) can heat the ejecta to super-virial temperatures, and the extremely high column density detected along these sightlines can be matched with a typical blast energy of $10^{51}$ erg, ejecta mass of $\sim 10 \, M_\odot$, and ambient density of $\sim 0.1$ cm$^{-3}$.  The ejecta of a core-collapse supernova is $\alpha$-enhanced and has super-solar composition, which naturally explains these findings. However, the covering fraction of such objects in the sky is $\lesssim 1 \%$ and therefore cannot account for super-virial gas along all the detected lines of sight. 

We further investigate whether the newly detected \mgxii K$\alpha$ line could arise from the ISM of the Galaxy, instead of the extended region. Following \citealt{Lara-DI2024b,Roy2025}, we looked for the \mgxii K$\alpha$ line in the XRB sightlines. Out of 27 XRBs, 4 sources exhibit a statistically significant ($>3\sigma$) detection of the \mgxii absorption line. The absorption lines in two of the four sightlines are broad, suggesting an origin in the immediate environment of the source XRB. Thus, only in 2 out of 27 sightlines, corresponding to a fraction of $\sim 7.5 \%$, the \mgxii K$\alpha$ may arise in the ISM. This small fraction, together with the \cite{Lara-DI2024b,Roy2025} results on other lines related to the hot gas, suggests that the super-virial phase likely resides in the extra-planar region. 



\section{Conclusion}
\label{sec:conclusion}

In this work, we investigated the presence of $z=0$ CGM absorption lines in the X-ray spectra of the Blazar PKS\,2155-304 using the grating data of \emph{Chandra} and \emph{XMM-Newton} and obtained the following results:

\begin{enumerate}

\item We detect Hydrogen-like and Helium-like ions of various elements, and for the first time, detect the \mgxii K$\alpha$ absorption line in the MW CGM (see Fig. \ref{fig:detection} and Table \ref{tab:all_instruments} and \ref{tab:detection}).  

\item For the first time, we detect four temperature phases using X-ray absorption lines along PKS\,2155-304 in such studies (see Figs. \ref{fig:illustration} and \ref{fig:temp_compare}). 
Our detection of highly ionized species like \mgxii and \sixiv confirms the existence of super‑virial hot gas ($\approx 5\times 10^7$ K) along this sightline (see Table \ref{tab:phase}). The temperatures of the warm-hot (virial) and warm (sub-virial) phases are $\approx 2\times 10^6$ and $\approx 2\times 10^5$ K, respectively. 
The $4^{\rm th}$ phase (cool phase; $T<1.7\times 10^5$ K) is detected using the column density ratios of \ov (non-detection) and \oiv (detection).  

\item The hot gas is found to have a super-solar composition of [Mg/O] and [Si/O] and is $\alpha$-enhanced, which is in line with the previous studies along other lines of sight. [C/O] and [Ne/O] are super-solar in the virial phase (see Fig. \ref{fig:non-solar-composition} and \ref{fig:alpha_abundance}).

\item We find that the low-ionization species are blue-shifted (v$_{\rm los}\approx -100$ km s$^{-1}$), while high-ionization species may be red-shifted, indicating a scenario with cool inflowing gas, quasi-static warm-hot gas, and hot outflowing gas along this sightline, which has never been found in previous studies (see Fig. \ref{fig:vel}).

\item Importantly, this marks the first such investigation conducted at southern Galactic latitudes—previous analyses were confined to northern directions, providing clear evidence of the widespread nature of X‑ray absorbing gas.

\item We find the presence of a temperature valley between the warm-hot and hot phases. A similar temperature valley is present between the sub-virial and cool phases (see Fig. \ref{fig:illustration}). This indicates the multi-peaked temperature distribution in the CGM of the MW. A large variation (an order of magnitude) in the temperature and equivalent H column density of the hot gas across different sightlines suggests the inhomogeneous nature of the hot gas.

\end{enumerate}

\section*{Acknowledgments}

MSB thanks Rahul Sharma, Manish Kumar, Kinjal Roy, and Shiv Sethi for the useful discussion. 
S.D. is grateful for the support provided by NASA through the NASA Hubble Fellowship grant HST-HF2-51551.001-A awarded by the Space Telescope Science Institute, which is operated by the Association of Universities for Research in Astronomy, Incorporated, under NASA contract NAS5-26555. 
SM is grateful for the support from the National Aeronautics and Space Administration (NASA) through Chandra Award Number AR0-23014X issued by the Chandra X-ray Center, which is operated by the Smithsonian Astrophysical Observatory on behalf of NASA under contract NAS8-03060. SM additionally acknowledges support from the NASA ADAP grant 80NSSC22K1121. 
YK acknowledges support from DGAPA-PAPIIT grant IN102023.
AG acknowledges support from NASA ADAP grants 80NSSC24K0626 and 80NSSC22K0480. AG also acknowledges support from NASA through Chandra Award Numbers GO3-24126X issued by the Chandra X-ray Center, which is operated by the Smithsonian Astrophysical Observatory on behalf of NASA under contract NAS8-03060. 

%



\facilities{\it{Chandra}, \it{XMM-Newton}}

\software{ \texttt{CLOUDY} \citep{Ferland2017}, \texttt{CIAO} \citep{Fruscione2006}, \texttt{SAS}, \texttt{XSPEC} \citep{Arnaud1999},  \texttt{NumPy} \citep{Harris2020}, \texttt{Matplotlib} \citep{Hunter2007}
}


\bibliography{reference}{}
\bibliographystyle{aasjournal}

\end{document}